\documentclass[aps,prx,twocolumn,superscriptaddress]{revtex4-1}
\usepackage[colorlinks,linkcolor=blue,urlcolor=blue,citecolor=blue,anchorcolor=blue]{hyperref}

\usepackage{amsmath, amssymb}
\usepackage{graphicx}
\usepackage{color}
\usepackage{hyperref}
\usepackage{braket}
\usepackage{bbold}
\usepackage{lipsum}
\usepackage{xcolor}
\usepackage{placeins}
\usepackage{dsfont}
\usepackage[normalem]{ulem}

\newcommand{\rhohat}{\hat{\rho}}
\newcommand{\ahat}{\hat{a}}
\newcommand{\adag}{\hat{a}^\dagger}
\newcommand{\Calpha}{|C_\alpha^\varphi\rangle}
\newcommand{\Calphi}[2]{|C_{#1}^{#2}\rangle}

\newcommand{\sigz}{\hat{\sigma}_z}
\newcommand{\sigm}{\hat{\sigma}_-}
\newcommand{\sigp}{\hat{\sigma}_+}

\newcommand{\DD}{\mathbb{D}}

\newcommand{\ketbra}[2]{|#1\rangle\langle#2|}

\begin{document}

\title{Preparing Schr\"odinger cat states in a microwave cavity using a neural network}

\author{Hector Hutin}
\affiliation{Ecole Normale Sup\'erieure de Lyon, CNRS, Laboratoire de Physique, 69342 Lyon, France}

\author{Pavlo Bilous}
\affiliation{Max Planck Institute for the Science of Light, Staudtstr. 2, 91058 Erlangen, Germany}

\author{Chengzhi Ye}
\affiliation{Ecole Normale Sup\'erieure de Lyon, CNRS, Laboratoire de Physique, 69342 Lyon, France}

\author{Sepideh Abdollahi}
\affiliation{Ecole Normale Sup\'erieure de Lyon, CNRS, Laboratoire de Physique, 69342 Lyon, France}

\author{Loris Cros}
\affiliation{Ecole Normale Sup\'erieure de Lyon, CNRS, Laboratoire de Physique, 69342 Lyon, France}

\author{Tom Dvir}
\affiliation{Quantum Machines Inc., Tel Aviv, Israel}

\author{Tirth Shah}
\thanks{Present address: Q.ANT GmbH, Handwerkstrasse 29, 70565 Stuttgart, Germany}
\affiliation{Max Planck Institute for the Science of Light, Staudtstr. 2, 91058 Erlangen, Germany}

\author{Yonatan Cohen}
\affiliation{Quantum Machines Inc., Tel Aviv, Israel}

\author{Audrey Bienfait}
\affiliation{Ecole Normale Sup\'erieure de Lyon, CNRS, Laboratoire de Physique, 69342 Lyon, France}

\author{Florian Marquardt}
\affiliation{Max Planck Institute for the Science of Light, Staudtstr. 2, 91058 Erlangen, Germany}
\affiliation{Department of Physics, Friedrich-Alexander-Universit\"at Erlangen-N\"urnberg, 91058 Erlangen, Germany}

\author{Benjamin Huard}
\affiliation{Ecole Normale Sup\'erieure de Lyon, CNRS, Laboratoire de Physique, 69342 Lyon, France}

\date{\today}

\begin{abstract}
Scaling up quantum computing devices requires solving ever more complex quantum control tasks.  Machine learning has been proposed as a promising approach to tackle the resulting challenges. However, experimental implementations are still scarce. In this work, we demonstrate experimentally a neural-network-based preparation of Schr\"odinger cat states in a cavity coupled dispersively to a qubit. We show that it is possible to teach a neural network to output optimized control pulses for a whole family of quantum states. After being trained in simulations, the network takes a description of the target quantum state as input and rapidly produces the pulse shape for the experiment, without any need for time-consuming additional optimization or retraining for different states. Our experimental results demonstrate more generally how deep neural networks and transfer learning can produce efficient simultaneous solutions to a range of quantum control tasks, which will benefit not only state preparation but also parametrized quantum gates.
\end{abstract}

\maketitle

Quantum information processing demands exquisite control over quantum systems. While direct optimization methods have demonstrated success in quantum control, they often necessitate access to gradient information~\cite{Khaneja2005,Krotov2018,Doria2011,Somloi1993}, and require to be run for each task. Much more flexibility is promised by the novel powerful tools emerging from the domain of machine learning and artificial intelligence~\cite{Carleo2019,Dawid2022,Krenn2023}. Model-free reinforcement learning considers the experiment or a simulation as a black box and trains the network on it~\cite{August2018,Zhang2018,An2019,Niu2019,Fosel2018,Mackeprang2020,Zhang2020,Sweke2021,Wang2020b,Borah2021a,Porotti2021,Sivak2021,Ye2024}. A few experiments have demonstrated this strategy for optimized qubit state readout and initialization~\cite{Reuer2023, vora2024}, generation of unitaries~\cite{Baum2021}, demonstrating dynamical decoupling~\cite{Peng2022}, improved measurements \cite{Cao2023}, Bose-Einstein condensate preparation~\cite{Guo2021}, quantum control in trapped ions ~\cite{Ai2022}, or optimized quantum error correction~\cite{Sivak2023}. Yet many experiments can be modeled completely with only a few \emph{a priori} undetermined or evolving parameters, employing a differentiable simulation. This can make it feasible to take gradients directly through simulations, the most efficient  version of model-based reinforcement learning~\cite{Zhang2018,Abdelhafez2020,Leung2017,porotti2023gradient}. Neural-network-based variants of this approach have yet to be demonstrated in experiments. In this domain, it has recently been shown theoretically that one might train a network once for a whole family of tasks and then use it to rapidly generate control sequences for any desired tasks selected from that family~\cite{SauvageMintert_PRL_2022}, making the whole process even more efficient, exploiting the general concept of transfer learning (learning for some tasks and benefiting from this for other tasks).
\begin{figure}[!]
\begin{center}
  \includegraphics[width=0.95\linewidth]{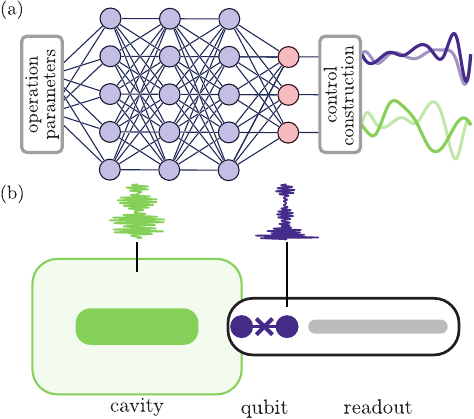}
      \caption{(a) A neural network takes the input parameters and outputs optimized control pulses characterizing a quantum operation.(b) These pulses drive a cavity (green) and its dispersively coupled qubit (purple) in order to prepare a desired quantum state. A readout resonator is used to measure the qubit state and perform Wigner tomography of the cavity.}
  \label{fig:setup}
 \end{center}
\end{figure}

In this work, we demonstrate model-based reinforcement learning for the preparation of quantum states in a cavity. We focus on arbitrary states within a specific class, namely two-component cat states in a cavity. This choice stems from the significant promise of such states for various quantum information processing applications such as quantum error correction and sensing~\cite{Cai2021}. We designed and trained a neural network to directly generate control pulses simultaneously applied to the cavity and a dispersively coupled qubit (Fig.~\ref{fig:setup}a). These control sequences are decomposed into a basis set of a few smooth basis functions, ensuring compatibility with existing real-time control hardware. We describe how to implement such a controller in an experiment and show how to estimate the preparation fidelity as fast as possible. We show that the network, once trained, is able to generate control sequences about five orders of magnitude faster than an established gradient-based technique. The rapidity of the control generation offered by the neural network could also prove instrumental in the case of real-time feedback protocols, needed in other tasks~\cite{Fosel2018,Mackeprang2020,Wang2020b,Borah2021a,Porotti2021,Sivak2021,Reuer2023,puviani2023boosting}.

A schematics of the experimental system is shown in Fig.~\ref{fig:setup}b (see also Appendix~\ref{sec:deviceandmeasurement}). The cavity whose state preparation is optimized by the neural network is the fundamental mode of a high-Q coaxial $\lambda/4$ resonator made of aluminum (green). Its frequency is $\omega_{\mathrm{c}}/2\pi=4.628~\mathrm{GHz}$ and its decay time is $T_{\mathrm{c}} = 225~\mathrm{\mu s}$. The cavity is dispersively coupled with a shift $\chi/2\pi = 238.5~\mathrm{kHz}$ to a transmon qubit (purple) resonating at $\omega_{\mathrm{q}}/2\pi=3.235~\mathrm{GHz}$, and whose decay time is $T_1 = 35$ $\mu$s and decoherence time $T_2$ varies between $20$ and $60~\mathrm{\mu s}$. The transmon qubit is also coupled to a readout mode composed of a $\lambda/2$ resonator connected to the readout line through a Purcell bandpass filter. Both the qubit and the cavity can be directly driven through distinct microwave ports.

In the interaction picture, the Hamiltonian of the qubit and cavity reads
\begin{align}\label{H}
&\hat{H}(t)/\hbar = -\chi \adag \ahat |1\rangle\langle 1|_{\mathrm{q}} +
\left[
\varepsilon_{\mathrm{c}}(t) \adag +  \varepsilon_{\mathrm{q}}(t) \sigp + \mathrm{h.c.}
\right]\;,
\end{align}
where $\varepsilon_{\mathrm{c}}(t)$ and $\varepsilon_{\mathrm{q}}(t)$ are the complex-valued time-dependent external fields driving the cavity and the qubit in their rotating frame, respectively. $\hat{a}$ is the annihilation operator of the cavity, $|1\rangle_{\mathrm{q}}$ is the first excited state of the transmon and $\sigp=\sigm^\dagger=|1\rangle\langle 0|_{\mathrm{q}}$ is the qubit raising operator. The system relaxation is included perturbatively in the simulations, as discussed in appendix~\ref{sec:relax_perturb}.

The preparation sequence is initiated when the system is in its ground state $|\psi(0)\rangle = \ket{0}_{\mathrm{q}} \otimes  \ket{0}_{\mathrm{c}}$. Both qubit and cavity are then driven during a fixed time $T$ in order to obtain a target state in the cavity. In this work we focus on target cat states of the form
\begin{equation}\label{eq:cat_def}
\Calpha\propto\ket{\alpha} + e^{-i\varphi} \ket{-\alpha}
\end{equation}
where $|\pm\alpha\rangle$ are coherent states of amplitude $\alpha$ and $\varphi$ is a phase. Additionally, we require the qubit to come back to the ground state, so that the full target state is
\begin{equation}
\ket{\mathcal{T}_\alpha^\varphi} = \ket{0}_{\mathrm{q}} \otimes  \Calpha_{\mathrm{c}}\;. 
\end{equation}

Under the action of $\hat{H}(t)$ in presence of decoherence, the density matrix of the system $\rhohat(t)$ evolves from $\rhohat(0) = \ketbra{\psi(0)}{\psi(0)}$ to $\rhohat(T)$. The fidelity to the target state is defined as $\mathcal{F}(\alpha,\varphi) = \bra{\mathcal{T}_\alpha^\varphi}\rhohat(T)\ket{\mathcal{T}_\alpha^\varphi}$. The optimization problem we set to solve consists in identifying the two complex (and thus four real) control functions $\varepsilon_{\mathrm{c}}(t)$ and $\varepsilon_{\mathrm{q}}(t)$ that maximize the fidelity $\mathcal{F}(\alpha,\varphi)$ after a driving time $T$ for every $(\alpha, \varphi)\in \mathcal{S} = [0, 4]\times[0, \pi]$.

\section{Neural Network\label{sec:neural_network}}

To this end, we employ a neural network (NN). The NN input consists of the real parameters $\alpha$ and $\varphi$ characterizing the target state $\ket{\mathcal{T}_\alpha^\varphi}$. The NN output is a set of 9 expansion coefficients for each of the 4 real control fields in a B-spline basis (see appendix~\ref{sec:bspln}). The encoding of the pulses is thus very light, which potentially enables much faster communication with the controller. As an example, loading one full pulse sequence of $2~\mathrm{\mu s}$ on our OPX necessitates $4\times 2000$ points, which takes approximately 450~ms, whereas loading only the $4\times 9$ corresponding parameters takes 24 ms, limited by the network communication time, which is already an order of magnitude faster.

\begin{figure}[h!]
\begin{center}
  \includegraphics[width=1.00\linewidth]{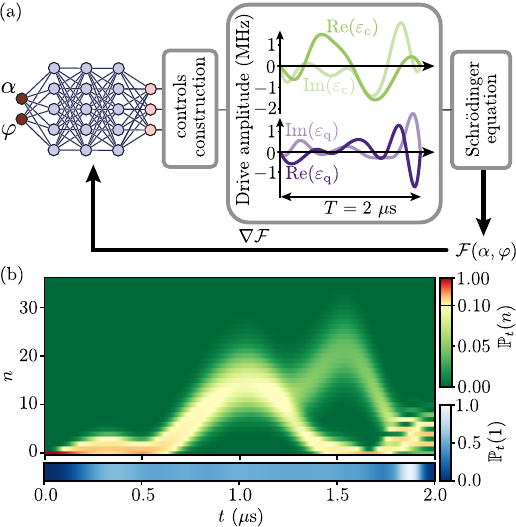}
  \caption{(a) Schematic representation of the data processing and training pipeline. Gradients can be taken through the network, control construction and simulation to update the network weights while optimizing the fidelity $\mathcal{F}$ via gradient ascent. The output of the network is a vector of 36 real numbers which is then converted into control pulses $\varepsilon_{\mathrm{c}}(t)$ and $\varepsilon_{\mathrm{q}}(t)$ with the goal of preparing a cat state $\Calpha\propto\ket{\alpha} + e^{-i\varphi} \ket{-\alpha}$ with $\varphi \in [0, 2\pi  ]$ and $\alpha \in [0, 4]$. (b) Simulated probability distribution of the photon number (top) and qubit excitation (bottom) as a function of time for the control fields shown in (a), which are generated by the neural network when $\alpha = 2$ and $\varphi = 0$. The fidelity of the predicted final state to $\Calpha$ is $\mathcal{F}(2, 0) = 94~\%$.}
  \label{fig:pipeline}
 \end{center}
\end{figure}

In order to apply the reinforcement learning approach, the NN is included in the data processing pipeline shown in Fig.~\ref{fig:pipeline}a. 
Each real control field is constructed by summing up the B-splines weighted by the output coefficients of the NN. In order to avoid the computational overhead associated with the implementation of a full density matrix $\rhohat(t)$ simulation, the coefficients are passed to a simpler Schr\"odinger equation solver, which computes the system state evolved under these control fields $\ket{\psi(\alpha, \varphi; t)}$. The fidelity $\mathcal{F}(\alpha, \varphi)$ can be approximated by
\begin{align}
    \mathcal{F}(\alpha, \varphi) \simeq \bigl| \braket{\mathcal{T}_\alpha^\varphi|\psi(\alpha, \varphi; t=T)} \bigr|^2 -\Delta F\;,
\end{align}
where $\Delta F$ is a first-order correction to the fidelity due to the decoherence ($T_1$, $T_2$, $T_{\mathrm{c}}$, see appendix~\ref{sec:relax_perturb}). 
We then use the infidelity as the loss function that has to be minimized:
\begin{equation}\label{eq:loss_def}
\mathcal{L}(\alpha, \varphi) = 1 - \mathcal{F}(\alpha, \varphi)\;.
\end{equation}
All these operations are differentiable, so gradients of the loss with respect to the NN parameters can be computed through the complete pipeline comprising the network, control construction and simulation, enabling thus gradient-based NN training.
The NN is trained on target cat states $\Calpha$ sampled randomly  within our parameter space $\mathcal{S}$. After a successful training, the NN is able to generate control fields which produce the cat state for \textit{any} $(\alpha, \varphi) \in \mathcal{S}$ in a preparation time $T = 2~\mathrm{\mu s}$, which was selected to be close to $\pi/\chi$.

The numerical modelling is performed on a time grid with 40 discretization intervals at the NN training stage and 200 intervals at the subsequent testing stage. The highest included Fock state $N_\mathrm{max}$ for the cavity (determining also the output size of the pipeline in Fig.~\ref{fig:pipeline}a) is varied during the NN training between $N_\mathrm{max}= 20$ and $N_\mathrm{max}=60$, and fixed to the value $N_\mathrm{max}=70$ at the testing stage. We made sure that this choice is sufficient for accurate modelling of all quantum states considered in this work. We use a NN of the usual dense architecture with 3 hidden layers of the size 30, 60 and 30 neurons, respectively (in total approx. 5000 trainable parameters). It is trained using the Adam algorithm~\cite{kingma2017adam} on batches sampled from ${\mathcal{S}}$. The batch size is gradually increased with the training stages. We use uniform sampling for $\varphi$, whereas $\alpha$'s are sampled with a probability linearly growing with larger values. This accounts for the fact that the NN training for the more interesting larger amplitudes $\alpha$ is more difficult. Additionally, at the NN training stage both $\alpha$ and $\varphi$ intervals are slightly extended beyond the edges in order to avoid the presence of boundary effects in ${\mathcal{S}}$. The pipeline discussed in the text above and shown in Fig.~\ref{fig:pipeline}a was implemented in Python using the JAX~\cite{jax2018github} and FLAX~\cite{flax2020github} libraries. We additionally benchmark our computations by performing the same system modelling with the QuTiP package~\cite{QuTiP, QuTiP2}, to extract fidelities. In appendices~\ref{sec:nn_training} and~\ref{sec:nn_performances} we provide further details on the NN training and performance study, respectively.

An example of the result of such simulations is shown in Fig.~\ref{fig:pipeline}b, where the evolution of the cavity photon number distribution and qubit excitation is plotted as a function of time when the neural network optimizes the creation of $|\mathcal{C}_2^0\rangle$. The qubit and cavity get quickly entangled leading to $\mathbb{P}_t(1_{\mathrm{q}})\approx 1/2$ for the qubit as the number of photons in the cavity rises. The cavity goes through states with large photon numbers and a superposition of low and large numbers arises around 1.6~$\mu\mathrm{s}$. At the end, the parity of the photon number seems to flip as well as the qubit state. Since large photon numbers are known to lead to dynamics that are not captured by a simple model~\cite{khezri2023, dumas2024}, it is desirable to keep them small. 
Here, no constraint was imposed on the maximal photon number during training, but the chosen B-spline basis with broad basis functions indirectly limits that number. Indeed finely controlling a system that evolves at a speed $\chi n$ in phase space requires pulses with a bandwidth of at least $\chi n$. The bandwidth of the B-spline scales inversely with their number. The chosen basis thus embeds a hidden cost naturally preventing the network from finding solutions with too high photon numbers. Tuning the number of elements in the basis then allows to control this limitation, that can be pushed to prepare cat states with a higher number of photons.

\section{Experimental results}

\begin{figure}[ht]
\begin{center}
  \includegraphics[width=1.00\linewidth]{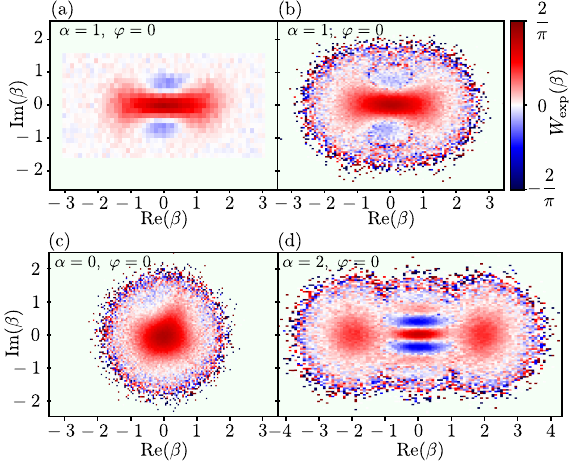}
  \caption{(a) Measured Wigner function $W_\mathrm{exp}(\beta)$ of the cavity for $\alpha = 1$ and $\varphi = 0$ and uniform sampling. Greed (b) Same preparation evaluated with optimal sampling. (c) and (d) Measured Wigner functions of the cavity with optimal sampling for $\alpha = 0$ and $2$, and $\varphi = 0$.}
  \label{fig:exp_scheme}
 \end{center}
\end{figure}

We now test the NN controller on the experiment for several targeted cat states $\Calpha$. After the cavity and the qubit are prepared in their ground states (see appendix~\ref{sec:heralding}), we apply the pulse sequence generated by the neural network on the cavity and qubit drive lines. A Wigner tomography is then performed using the same qubit in order to directly measure the Wigner function $W_\alpha^\varphi(\beta)$ of the cavity after the pulse sequence~\cite{Davidovich1996,Bertet2002}. It consists in mapping the displaced parity operator $\hat{\Pi}(\beta)=\hat{D}(\beta)e^{i\pi \hat{a}^\dagger\hat{a}}\hat{D}(-\beta)$ onto the qubit state before reading it out. The Wigner function is then given by $W_\alpha^\varphi(\beta)=2\langle\hat{\Pi}(\beta)\rangle/\pi$. We introduce $W_\mathrm{exp}(\beta)$, which is obtained by averaging the measurement outcomes of the observable $2\hat{\Pi}(\beta)/\pi$. In  Fig.~\ref{fig:exp_scheme}a, it is averaged over 1000 realizations per pixel for $\Calphi{1}{0}$.

From the Wigner function $W_\alpha^\varphi(\beta)$, we can compute the fidelity to the target cat state $\Calpha$ as (see appendix~\ref{sec:fidelity_estimation})
\begin{equation}
\mathcal{F}(\alpha, \varphi) = \pi \int_\mathbb{C} W_\alpha^\varphi(\beta)W_{\Calpha\langle \mathcal{C}_\alpha^\varphi|}(\beta)\mathrm{d}\beta
\end{equation}
with $W_{\Calpha\langle \mathcal{C}_\alpha^\varphi|}(\beta)=\frac{2}\pi\mathrm{Tr}(\hat{\Pi}(\beta)\Calpha \langle \mathcal{C}_\alpha^\varphi|)$  the Wigner function of the target state. Based on this formula, we can build an experimental estimator of $\mathcal{F}(\alpha, \varphi)$ (details in appendix~\ref{sec:fidelity_estimation}).

When performing state tomography of an unknown state, it makes sense to sample uniformly $W_\alpha^\varphi(\beta)$. However, it is possible to obtain a theoretically unbiased estimation of the preparation fidelity to a known target state with maximal precision using another sampling~\cite{Sivak2022} (in our case typically half the uncertainty compared to uniform sampling). In Fig.~\ref{fig:exp_scheme}b, we show the average measurement outcome $W_\mathrm{exp}(\beta) $ of $2\hat{\Pi}(\beta)/\pi$ for $N=10^6$ samples distributed in $\beta$ according to the optimal probability density $p_\mathrm{opt}(\beta) \propto|W_{\Calpha\langle \mathcal{C}_\alpha^\varphi|}(\beta)|$ for the target state $|\mathcal{C}_1^0\rangle$. Regions with low value for the target Wigner function appear very noisy owing to the small number of samples per pixel. For Fig.~\ref{fig:exp_scheme}b, one finds $\mathcal{F}(1, 0) = 94.7~\%\pm0.4~\%$. Details about this estimation and the errors introduced by the decay during Wigner tomography and by the double use (preparation and measurement) of the qubit can be found in appendix~\ref{sec:expt_implementation}. The estimated fidelities for two other target states in Fig.~\ref{fig:exp_scheme}c and d are $\mathcal{F}(0, 0) = 94.3~\%\pm0.2~\%$ and $\mathcal{F}(2, 0) = 91.1~\%\pm0.3~\%$.
We deliberately focused on training the NN for large values of $\alpha$, which explains the low fidelity when preparing the vacuum state.
\begin{figure}[h!]
\begin{center}
  \includegraphics[width=1.00\linewidth]{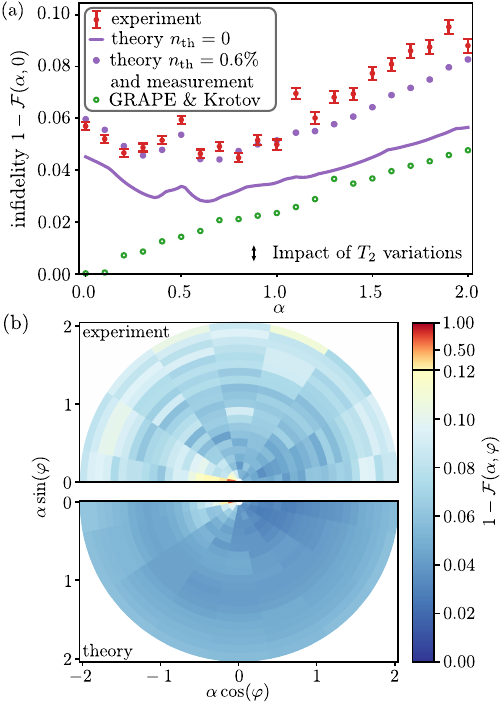}
\caption{(a) Red dots: measured infidelity $1-\mathcal{F}(\alpha,0)$ for 21 values of $\alpha$. Each point is obtained by averaging $10^6$ samples. Purple line: simulated infidelity $1-\mathcal{F}(\alpha,0)$ for the same parameters. Blue dots: infidelities obtained by simulating also the measurement procedure and taking into account the finite thermal population $n_\mathrm{th}$ of the cavity. Green open circles: simulated infidelity using pulses optimized by GRAPE followed by the Krotov method~\cite{QuTiP2,Khaneja2005,Goerz2019}. Double arrow: typical amount of infidelity change due to possible $T_2$ variations between 20~$\mu$s and 60~$\mu$s over the 8 days of measurement (see appendix~\ref{sec:effect_of_T2}). (b) Bottom: simulated infidelity for a sampling of the whole half-disk (reflected) in the parameter space $(\alpha\cos(\varphi), \alpha\sin(\varphi))$. Top: experimental infidelities obtained with $10^5$ shots sampled optimally. The measurement of these $20\times 9$ infidelities took about 8 days as well.}
  \label{fig:fidelities}
 \end{center}
\end{figure}

We use this measurement protocol to characterize the performance of the neural network in the preparation of many quantum states. In Fig.~\ref{fig:fidelities}a (red dots), we show the estimated preparation infidelities $1-\mathcal{F}(\alpha, \varphi)$ as a function of $\alpha$ for $\varphi = 0$ ($N=10^6$ samples per parameter). For comparison, we simulate the evolution of the density matrix $\hat{\rho}(\alpha,\varphi,t)$ under the same drives and compute predicted infidelities $1-\mathcal{F}(\alpha, \varphi)=1-\pi \int_\mathbb{C} W_{\hat{\rho}(\alpha,\varphi,T)}(\beta)W_{\Calpha\langle \mathcal{C}_\alpha^\varphi|}(\beta)\mathrm{d}\beta$ (purple line in Fig.~\ref{fig:fidelities}a). Both the experiment and the prediction exhibit the same qualitative behavior. For the smallest amplitudes, the infidelity decreases before increasing steadily. At low amplitudes, the small sampling (proportional to $\alpha$) of the network training dominates the infidelity of the preparation and it gets better as $\alpha$ rises. Around $\alpha\approx 0.5$, the infidelity grows with $\alpha$ because it is harder and harder to faithfully prepare a cat state with higher number of photons both theoretically and experimentally. We further find that dissipation dominates the theoretical infidelities: the same neural network on an idealized dissipation-free system achieves 2 orders of magnitude smaller infidelities.

In order to illustrate the trade-off between preparation infidelity and duration of pulse generation, we computed the infidelity one should obtain using a standard direct optimization method (see appendix~\ref{sec:GRAPE} for details). The green open circles in Fig.~\ref{fig:fidelities}a represent the theoretical infidelities obtained using $2~\mu\mathrm{s}$ long control pulses optimized by GRAPE  (GRadient Ascent Pulse Engineering) followed by the Krotov method \cite{QuTiP2,Khaneja2005,Goerz2019}. Note that in contrast to the 9 coefficients we use to parameterize the control pulses with the NN, GRAPE \& Krotov uses 2000 coefficients (one per ns). The infidelities are lower by about $0.01$ than with the NN except in the region of small $\alpha$ where the sampling in the NN training was scarce. However, generating the control pulses with GRAPE \& Krotov for the 21 values of $\alpha$ plotted in Fig.~\ref{fig:fidelities} a took about 6 hours on 14 cores i9 Intel CPU. This significantly exceeds the time of approximately 0.5 hours needed to train the NN on a GPU. More importantly, once trained, the NN can generate control signals for any cat state from the parameter space about five orders of magnitude faster than using GRAPE \& Krotov: generating the 21 control pulses takes $0.3$~s on an i9 Intel CPU. Note that the generation of pulses by the NN could be considerably accelerated using dedicated hardware (GPU, FPGA, TPU\ldots).

We also measure the preparation infidelities for several phases with $N=10^5$ samples per parameter pair $(\alpha, \varphi)\in[0.1, 2]\times [0, \pi]$ (Fig.~\ref{fig:fidelities}b upper panel). The predicted infidelities for the same parameter set are shown in the bottom panel of Fig.~\ref{fig:fidelities}b. From this plot, it appears that the neural network behaves similarly for any phase $\varphi$ except at low amplitudes $\alpha$. Beyond scarce sampling at low $\alpha$, this can be understood since the cat state varies rapidly with $\varphi$ next to $\pi$. At low amplitude $\alpha$, cat states $\Calpha$ are close to $|0\rangle$ except in the vicinity of $\varphi=\pi$ where $\Calpha$ is close to the Fock state $|1\rangle$. The large infidelity at $\varphi=\pi$ illustrates that a neural network performs poorly in the neighborhood of singularities since it inherently yields solutions which are continuous with respect to the input parameters.

Overall, we observe larger infidelities in the experiment than in the prediction. We attribute this mismatch to two main phenomena (see appendix~\ref{sec:errors}). The first one is an underestimation of the fidelity by the measurement protocol. Indeed the qubit is not always in the ground state at the end of the pulse sequence, which can artificially lower the measured fidelity, since the same qubit is then re-used for the Wigner tomography. Likewise, but with a lesser impact, the parity measurement procedure has a finite duration $t_\mathrm{par} = \frac{\pi}{\chi}$, during which the cavity decay probability is non negligible. The second reason for a deviation between experimental and theoretical fidelities is an imperfect state initialization. After the qubit and cavity reset, there is a remaining $n_\mathrm{th} = 0.6~\%$ thermal population in the cavity (see appendix~\ref{subsection:vacuum}), which is directly responsible for the same amount of fidelity decrease $\Delta \mathcal{F}_\mathrm{th} \simeq n_\mathrm{th}$. This can be understood simply: when the cavity (or the qubit) is in the Fock state $\ket{1}$ before the pulse sequence, the system is in an state that has zero overlap with the expected initial state (ground state). Thus, at the end of the unitary evolution, it will also have zero overlap with the expected final state, thus exhibiting zero fidelity to the target state. On average, the fidelity is then lowered by $\Delta \mathcal{F}_\mathrm{th} \simeq n_\mathrm{th}$. 
Simulating the whole measurement procedure and taking into account the remaining thermal population $n_\mathrm{th}$ of the cavity, we are able to capture the measured infidelities to a higher degree of accuracy (purple dots in Fig.~\ref{fig:fidelities}a).

The measured fidelities seem to vary a bit more than expected from the statistical error bars in Fig.~\ref{fig:fidelities}a, which is not the case of the theoretical predictions (purple solid line and purple dots). We explain this behavior by the variability of the qubit decoherence time $T_2$ during the experiment. While the value of $T_2$ for the theoretical predictions in Fig.~\ref{fig:fidelities}a was fixed to $42~\mathrm{\mu s}$, we have observed variations from 20~$\mu$s to 60~$\mu$s during the experiment. Since each of the two panels in Fig.~\ref{fig:fidelities} required 8 days of measurement records, it is likely that $T_2$ varied substantially from one point to the next. As shown in appendix~\ref{sec:effect_of_T2}, our measurement can be explained by variations of $T_2$ that are compatible with our observations.

\section{Conclusion}

In conclusion, we successfully trained a neural network to output controls that prepare a wide range of quantum states with good fidelity. We showed that the comparatively simple neural network, once trained, generates new control sequences about five orders of magnitude faster than GRAPE followed by the Krotov method, at the expense of only a 0.01 drop in fidelity for the states of interest. In the future, applying such neural networks may be key for potential applications requiring fast change of pulse parameters (feedback, adaptive techniques, error correction etc.).  There are multiple potential follow-up applications: one may promote the model parameters as trainable, and directly use the experimental measurements as input to learn the correct model parameters from the experiment. Alternatively, one may train for a whole range of model parameters in advance and adapt them to the experiment on the fly, e.g. when parameters are drifting. Since the network structure is light and easy to handle, it can form an efficient building block for complex network-based real-time feedback control of quantum systems. In that setting the moderate size of the network, efficient communication and information compression are crucial for low-latency data processing, whether the network is hosted on the FPGA controller itself or on a dedicated hardware.

\begin{acknowledgments}
This research was supported by the QuantERA grant ARTEMIS, by ANR under the grant ANR-22-QUA1-0004 and by German Federal Ministry of Education and Research under the grant 13N16360 within the program ``from basic research to market''. We also thank the Munich Quantum Valley, which is supported by the Bavarian state government with funds from the Hightech Agenda Bayern Plus. We acknowledge IARPA and Lincoln Labs for providing a Josephson Traveling-Wave Parametric Amplifier. We thank Antoine Essig, Omar Fawzi, Thomas F\"osel, Cl\'ement Panais, and Arthur Strauss for useful discussions.
\end{acknowledgments}

\bibliographystyle{apsrev4-1}

\appendix

\section{B-spline basis set}\label{sec:bspln}

Instead of working with the control fields directly, we consider their expansion in a B-spline basis in the form described in Ref.~\cite{Johnson_book_2007} in the context of computational atomic physics applications. A crucial advantage of such an approach is a possibility to pre-load the B-splines on the hardware and send only the expansion coefficients of control signals for different cat states. A particular B-spline basis is built on a fixed knot sequence $t_1 \le t_2 \le \ldots \le t_{n+k+1}$ and is characterized by the basis size $n$ and the degree $k$ of the B-splines.  Each $k$-degree B-spline $B_{i, k}(t)$ with $i=1,  \ldots, n$ is a piecewise polynomial of degree $k$ inside the interval $[t_i, t_{i+k+1})$ and vanishes outside this interval. They are constructed recursively as follows:
\begin{equation}
B_{i, 0}(t) =
\begin{cases}
    1, 			& \text{if } t \in [t_{i},  t_{i+1})  \;, \\
    0,           & \text{otherwise}\;,
\end{cases}
\end{equation}
\begin{equation}
B_{i, k}(t) = \frac{t - t_i}{t_{i+k}-t_i} B_{i, k-1}(t) +  \frac{t_{i+k+1}-t}{t_{i+k+1}-t_{i+1}} B_{i+1, k-1}(t)\;.
\end{equation}
We follow Ref.~\cite{Johnson_book_2007} and choose for the $k + 1$ leftmost knots $t_1 = \ldots = t_{k+1} = 0$ and for the $k + 1$ rightmost knots $t_{n+1} = \ldots = t_{n+k + 1} = T$.  The rest $n - k - 1$ knots are distributed uniformly between 0 and $T$.

\begin{figure}
\begin{center}
  \includegraphics[width=\linewidth]{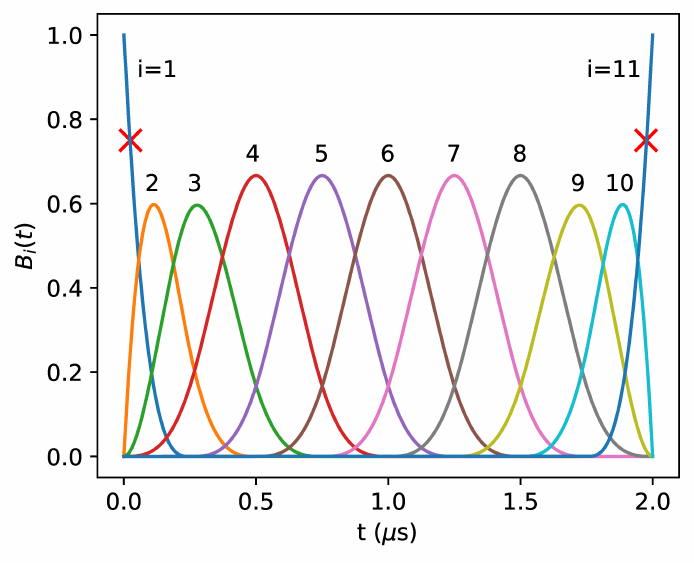}
  \caption{B-spline basis of size $n=11$ and degree $k=3$. In this work the first and last B-splines are excluded in order to restrict the driving signals to those starting and ending at zero.}
  \label{fig:bspln}
 \end{center}
\end{figure}

In this work we stick to a B-spline basis set of size $n=11$ and degree $k=3$ shown in Fig.~\ref{fig:bspln} for the excitation time interval $T=2\;\mu\mathrm{s}$. We exclude the first and last B-splines which are the only ones having non-zero values at the interval edges. In this way, we restrict our driving signals to start and end with zero amplitude. The problem of the control field optimization reduces to searching for a total of $9 \times 4$ optimal expansion coefficients for the 2 complex (qubit + cavity), and thus 4 real driving fields. Apart from a compact representation of the control signals, B-splines offer also a practical advantage of being non-zero only in restricted intervals. In particular, for the chosen basis set only up to 4 B-splines are non-zero at each point of the driving time interval. This ensures compatibility with existing real-time quantum control hardware. For instance, the OPX from Quantum Machines can generate up to 18 signals at the same time, which would allow to use 4 B-splines for 4 control fields at the same time.

\section{Neural network training\label{sec:nn_training}}

Here we show details of the neural network training performed in the way described in Section~\ref{sec:neural_network}. Whereas we experimentally test preparation of cat states with parameters $\alpha \in (0, 2)$ and  $\varphi \in (0, \pi)$, the neural network training is performed on a wider and more challenging parameter space $\alpha \in (-4, 4)$ and  $\varphi \in (0, \pi)$. Due to the equality $\Calphi{\alpha}{2\pi - \varphi} = e^{i \varphi} \Calphi{-\alpha}{\varphi}$ obtained directly from the definition of cat states (\ref{eq:cat_def}), the neural network is equivalently able to predict control signals for $\alpha \in (0, 4)$ and  $\varphi \in (0, 2\pi)$.

\begin{figure}[h!]
\begin{center}
  \includegraphics[width=1.00\linewidth]{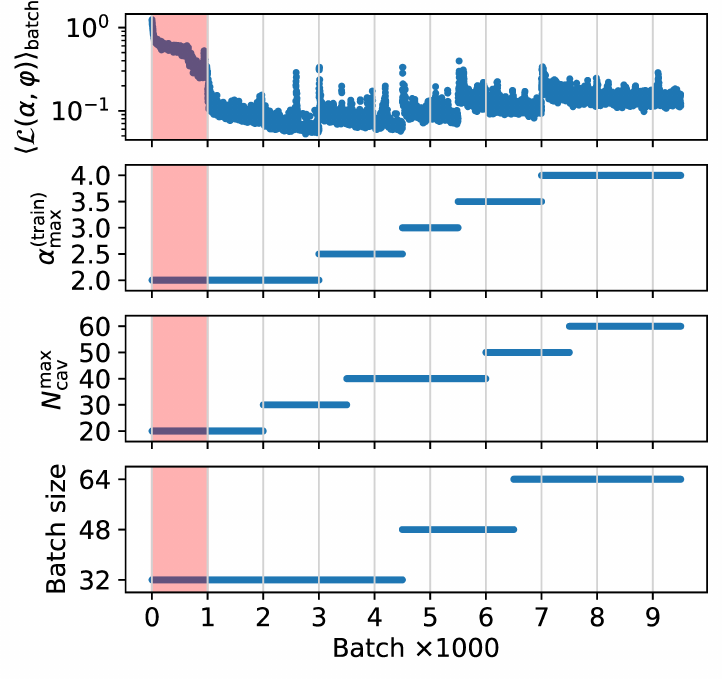}
  \caption{Neural network training progress with batches. In the upper panel we show the batch-averaged loss $\left<\mathcal{L}(\alpha, \varphi)\right>_\mathrm{batch}$, whereas the other panels show the gradual change of the parameters $\alpha^\mathrm{(train)}_\mathrm{max}$, $N^\mathrm{max}_\mathrm{cav}$, and the batch size. See text for details.}
  \label{fig:nn_training}
 \end{center}
\end{figure}

In Fig.~\ref{fig:nn_training} we demonstrate the training procedure performed in a few stages by gradually increasing the ``complexity'' of the parameter space and the numerical model of the system. Concretely:
\begin{itemize}
\item we increase the $\alpha$-interval $(0, \alpha^\mathrm{(train)}_\mathrm{max})$ from $\alpha^\mathrm{(train)}_\mathrm{max}=2$ to $\alpha^\mathrm{(train)}_\mathrm{max}=4$;
\item the number of Fock states included in the cavity numerical modeling is increased from $N^\mathrm{max}_\mathrm{cav}=20$ to $N^\mathrm{max}_\mathrm{cav}=60$;
\item since the volatility of the training loss is larger for a larger parameter space, we increase also the batch size at later training stages.
\end{itemize}

At the starting stage (first 1000 batches) shown by the red-shaded area, the training often stagnates and the batch-averaged loss $\left<\mathcal{L}(\alpha, \varphi)\right>_\mathrm{batch}$ does not decrease, see the upper panel of Fig.~\ref{fig:nn_training}. This is a local minimum in which the neural network learns not to drive the qubit at all, whereas the cavity remains in a coherent state. We overcome this trap by adding an incoherent excitation process from ground to excited state with a rate $\Gamma_{\uparrow}=(5~\mu\mathrm{s})^{-1}$ at this initial stage. We stress that this pumping is not present in the experiment and is used here only to tackle the mentioned training problem. Numerically this is achieved within the perturbative framework for inclusion of relaxation channels described in Appendix~\ref{sec:relax_perturb}. While such a trapping occurred at the beginning, training proceeded smoothly afterwards.

\section{Neural network performance}\label{sec:nn_performances}

\begin{figure}[h!]
\begin{center}
  \includegraphics[width=1.00\linewidth]{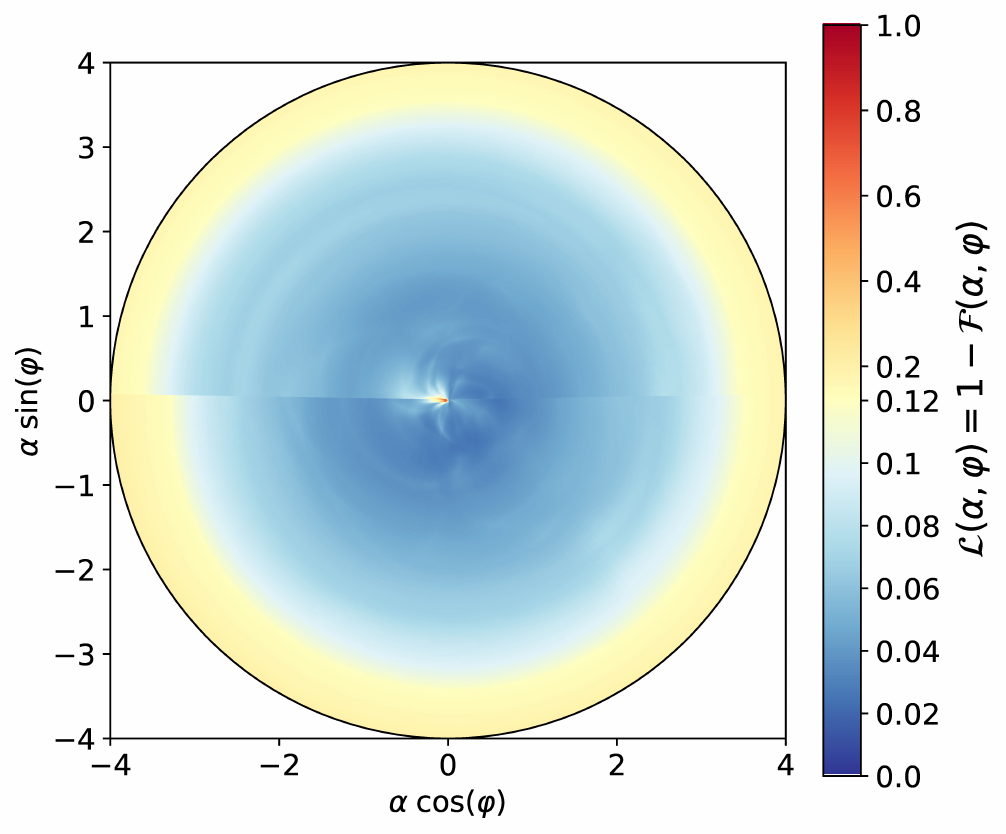}
  \caption{Simulated infidelity of cat states obtained by driving the system in the way suggested by the trained neural network. The parameter space $(\alpha, \varphi)$ is shown in polar coordinates. The color indicates the infidelity $\mathcal{L}(\alpha, \varphi) = 1 - \mathcal{F}(\alpha, \varphi)$.}
  \label{fig:cats_circ}
 \end{center}
\end{figure}

In order to check the neural network performance, we apply it first to a selection of size 3200 from the parameter space $\alpha \in (0, 4)$ and  $\varphi \in (0, 2\pi)$. In Fig.~\ref{fig:cats_circ} we show the simulated fidelity of the target states obtained by driving the system as suggested by the neural network. The parameters $(\alpha, \varphi)$ are encoded in polar coordinates. We observe indeed that the obtained fidelity is worse for larger $\alpha$-s justifying sampling with linearly growing density. We checked that the final fidelity for states with $\alpha > 2$ (which were not experimentally tested in this work) can be improved by increasing the number of the B-splines in the basis set.

\section{Accounting for decoherence\label{sec:relax_perturb}}

In general, the decoherence effects cannot be addressed using the Schr\"odinger equation, and need switching to an open quantum system approach such as the Lindblad equation for the density matrix. In this work, however, we are not interested in the whole system dynamics upon inclusion of decoherence, but only in the correction of the final state fidelity. Therefore, instead of solving the more complicated Lindblad equation, we still solve the Schr\"odinger equation and correct the fidelity for the decoherence using first-order perturbation theory.

We derive here the aforementioned correction. The Lindblad equation reads
\begin{equation}\label{eq:lindblad}
\frac{\partial \hat{\rho}}{\partial t} = -i [\hat{H}, \hat{\rho}] + \sum_i \frac{1}{\tau_i} D_i[\hat{\rho}]\;,
\end{equation}
where the summation is performed over all decoherence channels with respective characteristic times $\tau_i$. For each decoherence channel
\begin{equation}
D_i[\hat{\rho}] = \hat{a}_i \hat{\rho} \hat{a}_i^\dagger - \frac{1}{2}\hat{a}_i^\dagger \hat{a}_i \hat{\rho} -  \frac{1}{2}\hat{\rho} \hat{a}_i^\dagger \hat{a}_i \;,
\end{equation}
where $\hat{a}_i$ is the corresponding jump operator. In our case, for instance, the following decoherence channels (governed by the jump operators shown in the parenthesis) are present:
\begin{itemize}
    \item cavity dissipation (photon annihilation operator $\hat{a}$);
    \item qubit dissipation ($\hat{\sigma}_-$);
    \item qubit dephasing ($\hat{\sigma}_z / \sqrt{2}$).
\end{itemize}

The density matrix at time $t$ in absence of decoherence is the pure state density matrix
\begin{equation}
\hat{\rho}^{(0)}(t) = \ket{\psi(t)}  \bra{\psi(t)}\;.
\end{equation}
The first-order correction due to the $i$-th decoherence channel is
\begin{equation}
\hat{\rho}^{(1)}_i(t) = \frac{1}{\tau_i} \int_0^t dt' \; \hat{U}^{(0)}(t', t) \; D_i[\hat{\rho}^{(0)}(t)] \;\hat{U}^{(0)\dagger}(t', t)
\end{equation}
leading to the change in the final state fidelity at $t=T$ with respect to the target state $\ket{\mathcal{T}}$
\begin{equation}
\Delta F_i = \braket{ \mathcal{T} | \; \hat{\rho}^{(1)}_i(t=T) \;|\mathcal{T} }\;,
\end{equation}
We make at this point an additional assumption that the obtained state without decoherence $\psi(t=T)$ is very close to the target state $\ket{\mathcal{T}}$ and write
\begin{equation}
\Delta F_i \approx \braket{ \psi(t=T) | \; \hat{\rho}^{(1)}_i(t=T) \;| \psi(t=T) }\;.
\end{equation}
In this approximation, the evolution operator $\hat{U}(t', T)$ connects the intermediate system state at time $t'$ and the target system state. From here we obtain finally:
\begin{equation}\label{eq:dfid}
\Delta F_i  \approx \frac{1}{\tau_i}
\int_0^T dt'
\left[
\left| \braket{\psi(t')|\hat{a}_i|\psi(t')} \right|^2 -  \braket{\psi(t')|\hat{a}_i^\dagger \hat{a}_i|\psi(t')}
\right]
\;.
\end{equation}

\begin{figure*}[ht]
\centering
\includegraphics[width=\linewidth]{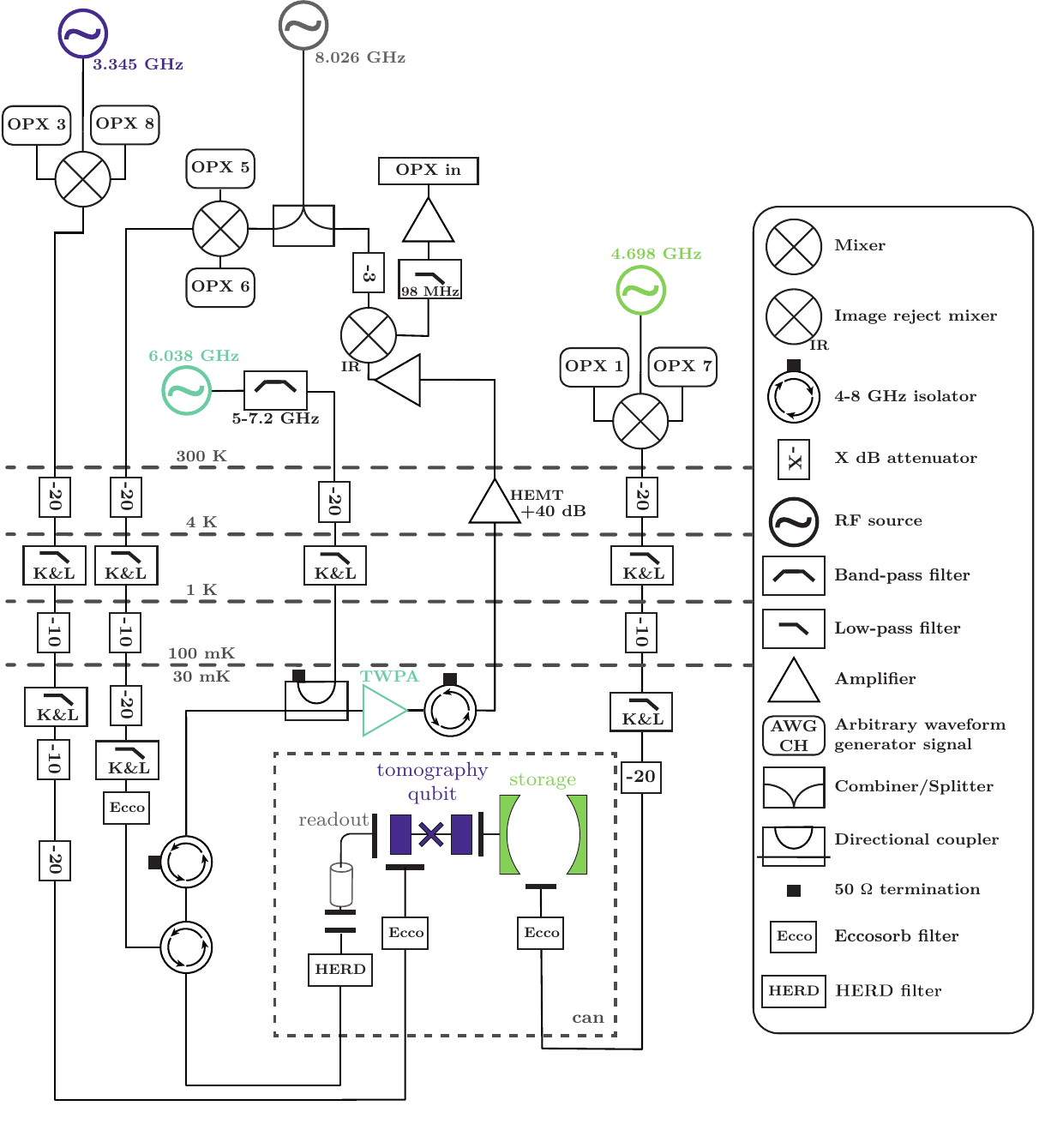}
\caption{Schematic of the measurement setup. A color matched RF source is dedicated to each element. Room-temperature isolators are not represented for the sake of clarity.}
\label{fig:cablage}
\end{figure*}

\begin{table*}\caption{Table of circuit parameters}
\begin{tabular}{ |p{6cm}||p{3cm}|p{3cm}|p{3cm}|  } 

 \hline
 \multicolumn{4}{|c|}{Table of circuit parameters} \\
 \hline
  \hline
 Circuit parameter& Symbol & Hamiltonian term & Value \\
 \hline
 Cavity frequency   & $\omega_{\mathrm{c}}/2\pi$   &$\hbar \omega_{\mathrm{c}}\adag\ahat$ &   $4.628~\mathrm{GHz}$\\
 Qubit frequency&   $\omega_{\mathrm{q}}/2\pi$  & $\hbar \omega_{\mathrm{q}} |1\rangle\langle 1|_{\mathrm{q}} $  & $3.235~\mathrm{GHz}$\\
 Readout frequency    & $\omega_r/2\pi$ & $\hbar \omega_r\hat{r}^\dagger \hat{r}$ & $7.960~\mathrm{GHz}$ \\
 Cavity-qubit cross Kerr rate    & $\chi/2\pi$ & $-\hbar \chi \adag\ahat |1\rangle\langle 1|_{\mathrm{q}}$ & $238.5~\mathrm{kHz}$ \\
 \hline
  Circuit parameter& Symbol & Dissipation operator & Value \\
 \hline
 Qubit decay time & $T_{\mathrm{q}}$  & $1/T_{\mathrm{q}} \DD_{\sigm}$   & 35 $\mathrm{\mu s}$ \\
 Cavity decay time & $T_{c}$  & $1/T_{c} \DD_{\ahat}$   & 225 $\mu$s \\
 Readout decay time & $T_{1, r}$  & $1/T_{1, r}\DD_{\hat{r}}$   &  80 ns \\
 Qubit dephasing time & $T_{q, \varphi}$  & $1/2T_{q, \varphi}\DD_{\sigz}$   & 175 $\mu$s \\
\hline
\end{tabular}
\label{circuits parameters}
\end{table*}

\section{Device and measurement setup}
\label{sec:deviceandmeasurement}
\subsection{Device fabrication}
\begin{figure}[h]
\centering
\includegraphics[width=\linewidth]{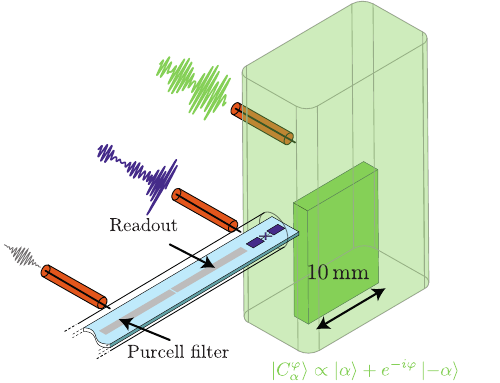}
\caption{Scheme of the device.}
\label{fig:cav_scheme}
\end{figure}

The system is composed of one 3D $\lambda/4$ coaxial cavity resonator in 99.99~\% pure Aluminum, into which a chip, containing the transmon qubit with its readout resonator and Purcell filter, is inserted (see Fig.~\ref{fig:cav_scheme}). This chip is made of an etched 200 nm thick film of sputtered Tantalum on a 430 $\mathrm{\mu m}$ thick sapphire substrate (deposited by Star Cryoelectronics, Santa Fe, USA). The Josephson junctions of both transmons are standard Dolan bridge e-beam evaporated Al/AlOx/Al junctions~\cite{thesisessig2021}.

\subsection{Measurement setup}

The readout resonator, the cavity and the qubit are driven on resonance by pulses that are generated using an OPX from Quantum Machines®. It has a sampling rate of 1~GS/s. The generated pulses are modulated at a frequency 68~MHz for readout, $\omega_{\mathrm{q}}^\mathrm{IF}/2\pi = 110$~MHz for the qubit, $\omega_{\mathrm{c}}^\mathrm{IF}/2\pi = 70$~MHz for cavity. They are up-converted using I-Q mixers with continuous microwave tones produced by three channels of an AnaPico® APUASYN20-4 for the readout resonator and cavity, Agilent® E8257D for the qubit. 

The reflected signals from the readout is amplified with a TWPA provided by Lincoln Labs~\cite{Macklin2015} powered by a pump tone at 6.038~GHz. The follow-up amplification is performed by a HEMT amplifier from Low Noise Factory® at 4~K and by room-temperature amplifiers. The signal is down-converted using image reject mixers before digitization by the input ports of the OPX.

The samples are placed inside a can comprising a layer of lead, gold-plated copper and a layer of cryoperm. The inside of the can is coated with an absorptive mixture of 86~\% Stycast 2850 FT, 7~\% catalyst 23 LV and 7~\% carbon powder.

For the cavity, the controls pipeline is as follows. We define $I_{\mathrm{c}}(t)$ and $Q_{\mathrm{c}}(t)$ the time dependent voltages to send on resonance to the cavity. These controls are then converted into $I_{\mathrm{c}}^\mathrm{IF}(t) = \cos(\omega^\mathrm{IF}_{\mathrm{c}}t)I_{\mathrm{c}}(t) - \sin(\omega^\mathrm{IF}_{\mathrm{c}}t)Q_{\mathrm{c}}(t)$ and  $Q_{\mathrm{c}}^\mathrm{IF}(t) = \sin(\omega^\mathrm{IF}_{\mathrm{c}}t)I_{\mathrm{c}}(t) + \cos(\omega^\mathrm{IF}_{\mathrm{c}}t)Q_{\mathrm{c}}(t)$ and sent on DAC OPX 5 and 6. These signals are then mixed with a local oscillator at frequency $\omega_{\mathrm{c}}^\mathrm{LO}$ with an IQ mixer in a lower sideband setting to give a RF voltage $V^\mathrm{RF}(t) \propto \cos{\big((\omega_{\mathrm{c}}^\mathrm{LO} -\omega^\mathrm{IF}_{\mathrm{c}})t\big)}I_{\mathrm{c}}(t) +\sin{\big((\omega_{\mathrm{c}}^\mathrm{LO} -\omega^\mathrm{IF}_{\mathrm{c}})t\big)}Q_{\mathrm{c}}(t) $ at the output of the IQ mixer. The pulsations $\omega_{\mathrm{c}}^\mathrm{LO}$ and $\omega^\mathrm{IF}_{\mathrm{c}}$ are chosen such that $\omega_{\mathrm{c}}^\mathrm{LO}-\omega^\mathrm{IF}_{\mathrm{c}} = \omega_{\mathrm{c}}$. Through attenuation in the lines and the coupling of the qubit to these lines, this translates to the control amplitudes $\varepsilon(t) = \xi_{\mathrm{c}}(I_{\mathrm{c}}(t)+iQ_{\mathrm{c}}(t))$ on the cavity. Details about the calibration of $\xi_{\mathrm{c}}$ can be found in~\ref{pulse calibration}. The qubit is driven similarly with control voltages $I_{\mathrm{q}}(t)$ and $Q_{\mathrm{q}}(t)$.

\section{Calibration and heralding}
\label{sec:heralding}
This section aims at giving technical details about the system initialization and measurement. The initialization part was especially complex, as we needed to make sure that both cavity and qubit are in the ground state. This is done using standard measurement-based feedback techniques. On top of this standard reset protocol, the presence of a Two-Level System (TLS) made the frequency of the qubit jitter between two frequencies at long timescales, which obviously deteriorates the fidelity of the state preparation if not mitigated. Here, we study this TLS dynamics and desmonstrate a heralding protocol for the qubit Fig.~\ref{fig:res_ramsey}, demonstrating close to total mitigation of the TLS detrimental impact.

\subsection{Qubit readout and reset}

\begin{figure}[h!]
\begin{center}
  \includegraphics[width=\linewidth]{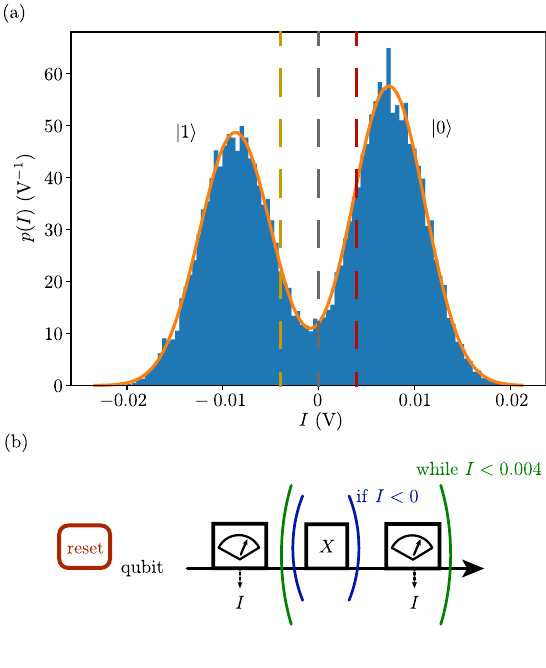}
  \caption{(a) Blue: Histogram of 20 000 single shot readouts of the qubit in its thermal state combined with 20 000 more after an approximate $\pi$ pulse. Dashed lines represent thresholds for qubit and cavity reset operations (see text). (b) Reset scheme. The qubit is read until the measurement record exceeds the threshold $I=0.004~\mathrm{V}$. If the qubit is more likely to be in the excited state, a $\pi$ pulse is applied before the readout.} 
  \label{fig:RO}
 \end{center}
\end{figure}

We readout the qubit with a square pulse of length $2.2~\mathrm{\mu s}$ and measure the quadrature $I$ that encodes the qubit information. Single-shot fidelity is around $98~\%$. The reset of the qubit consists in a feedback loop using two thresholds represented in Fig.~\ref{fig:RO}a. Exceeding the first one (red dashed line) is the ending condition, allowing to herald the qubit in its ground state $\ket{0}$ with more than $99.9~\%$ fidelity. The second threshold (gray dashed line) triggers a $\pi$ pulse (gate $X$) if the qubit is more likely to be in the excited state, (\emph{i.e} $I$ below threshold). Getting below a third threshold (yellow dashed line in Fig.~\ref{fig:RO}a) allows to herald the qubit in the excited state with the same fidelity, and is used for the reset of the cavity (see section \ref{subsection:vacuum}).

\subsection{TLS mitigation}

\begin{figure}[h!]
\begin{center}
  \includegraphics[width=1.0\linewidth]{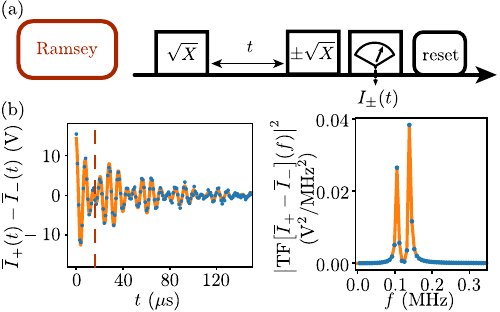}
  \caption{(a) Ramsey pulse sequence. The $\pi/2$ ($\sqrt{X}$) pulses are detuned from the qubit frequency by around 140~kHz, and are separated by a waiting time $t$. The rotation axis of the second pulse is reversed at each repetition. (b) Blue dots: Signal averaged over a 2000 repetitions as a function of $t$. Orange solid line: fit with the sum of two oscillations detuned by $\delta/2\pi = 32~\mathrm{kHz}$ and decaying with a time $T_2=51~\mu\mathrm{s}$. (c) Blue dots: Spectral density obtained by a Fourier transform of (b). Solid orange line: Spectral density of the fit function in (b). } 
  \label{fig:ramsey}
 \end{center}
\end{figure}

Probing the frequency of the qubit is done with a standard Ramsey pulse sequence. The signal is obtained by subtracting the results of two Ramsey sequences performed with opposite parity for the second $\pi/2$ pulse (see Fig.\ref{fig:ramsey}a). We see that the qubit has two possible frequencies (see Fig.~\ref{fig:ramsey}b). We attribute this to the presence of a spurious Two Level System (TLS) dispersively coupled to the qubit, with a dispersive shift $\delta/2\pi = 32~\mathrm{kHz}$. This TLS has a thermal occupation of about 40~\%, and its state switches over the course of the full experiment, which makes the frequency of the qubit jitter between $\omega_{\mathrm{q}}/2\pi$ and $\omega_{\mathrm{q}}/2\pi-\delta/2\pi$. The experimental temporal Ramsey signal fits well with a model including two frequencies with the same decay time $T_2$, which drifted between 20 and 60~$\mu$s over a few hours (see Fig.~\ref{fig:ramsey}b).

\begin{figure*}[h!]
\begin{center}
  \includegraphics[width=0.9\linewidth]{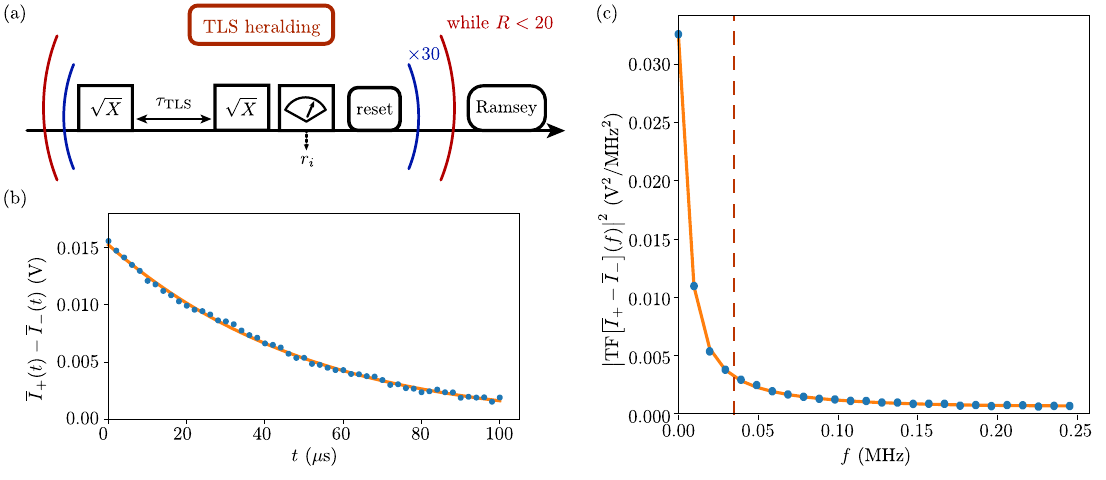}
  \caption{(a) Scheme of the TLS heralding procedure. (b) On-resonance Ramsey measurement performed using the TLS heralding procedure. Blue dot: experimental signal. Orange solid line: exponential fit with $T_2 = 50~\mu s$. (c) Blue dot and orange solid line: spectral density of (b) transform. Vertical dashed line: position of the expected second peak if it were not suppressed by the heralding procedure.}
  \label{fig:res_ramsey}
 \end{center}
\end{figure*}

The frequency jumps introduced by this TLS are detrimental to the fidelity of the prepared state, as the state preparation needs a precise calibration of the frequency of the qubit. We mitigated this effect using a heralding procedure. Since the TLS state cannot be inferred from the qubit state in one shot, we chose to take a decision using a repetition of 30 times the same Ramsey sequence on resonance with the frequency that the qubit takes when the TLS is in the ground state (identified by the highest peak in Fig.~\ref{fig:ramsey}), with a delay $\tau_\mathrm{TLS}=\frac{\pi}{\delta} = 15.5 ~\mathrm{\mu}s$. The pulse sequence is schematized in Fig.~\ref{fig:res_ramsey}a. To each of the 30 measurement outcomes $\{I_i\}_{1\leq i\leq 30}$, we associate
\begin{align}
r_i= \left\{
    \begin{aligned}\label{record}
        &1 \mathrm{~if~I_i<0~V}\\
        &0 \mathrm{~~else.} 
    \end{aligned}
\right.
\end{align}
From this, we compute $R = \sum_{i=1}^{30}r_i$. The heralding condition is then fixed to $R\geq 20$. The Figure \ref{fig:res_ramsey}a shows the pulse sequence used to perform a standard Ramsey measurement after this heralding sequence.  Figures \ref{fig:res_ramsey}b and c show the resulting decaying Ramsey signal and its Fourier transform. The beating in Fig.~\ref{fig:ramsey}b  disappears when using the heralding, as well as the second peak in the spectrum of Fig.~\ref{fig:ramsey}c, whose expected position is materialized by the vertical red dashed line in \ref{fig:res_ramsey}c. This demonstrates a successful mitigation of the TLS effects.

Further analysis of the TLS dynamics, and details about the determination of the heralding procedure is shown in the following.
\newline

\label{sec:thresholding}

\subsubsection{Dynamics of the TLS and high fidelity detection of its state}

We first studied the dynamics of the frequency jumps of the qubit, by performing $10^5$ times the same Ramsey measurement on resonance with the upper frequency in~\ref{fig:ramsey}b with a delay $\tau_\mathrm{TLS}$. The procedure is the same as for the parity measurement on a dispersively coupled harmonic oscillator \cite{Vlastakis2013}. The pulse sequence is shown in Fig.~\ref{fig:ramsey_corr}a. Without qubit decoherence, this would allow us to read out the state of the TLS in a single shot.

With the associated measurement signals $\{I_i\}$, we could first investigate the switching time of this TLS.

\onecolumngrid

\twocolumngrid

\noindent  To do this, we computed the correlations between the measurement records $I_i$ and $I_{i+k}$. We define $c_k = \mathrm{Cov}(\{I_{i}\}, \{I_{i+k}\})$, and plot this quantity as a function of $k$ (Fig.~\ref{fig:ramsey_corr}b). We observe a first decrease at short times $k<100$, corresponding to a correlation time shorter than 1 ms. This we attributed to the cavity thermal population. The much longer drift that follows is attributed to the frequency jumps. It is not clear to us why this does not really follow an exponential decay. We believe that the correlations at long times of the TLS state cannot be properly captured by our measurement because it is too short ($10^5$ shots). Fitting the beginning of the drifts gives a characteristic time of around 400 ms, which gives plenty of time to perform post-selection. In order to have a good fidelity, we need to accumulate statistics over several measurements, as $\tau_\mathrm{TLS}$ cannot be neglected compared to the decoherence time $T_2$ of the qubit. This measurement cannot be single shot.
 
\subsubsection{Heralding on the TLS state}

\begin{figure}[h!]
\begin{center}
  \includegraphics[width=1.0\linewidth]{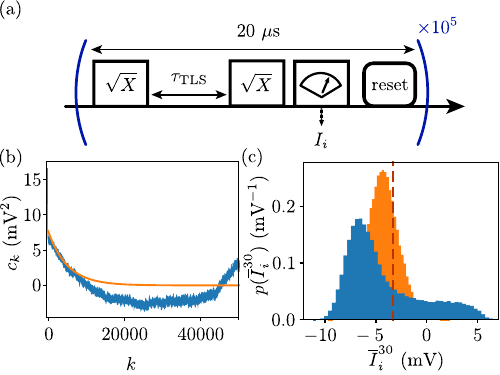}
  \caption{(a) Pulse sequence of the TLS state measurement. (b) Blue: Measured covariance $c_k$ as a function of $k$. Orange: Exponential fit for $k<10000$. The characteristic fitted time is $400~\mathrm{ms}$. (c) Blue: Probability density of the running average of the $\{I_i\}$. Orange: Same quantity, but with over $\{I_{\sigma(i)}\}$, where $\sigma$ is a random permutation. Red dashed line: Threshold used for Fig.~\ref{fig:selection_TLS}.} 
  \label{fig:ramsey_corr}
 \end{center}
\end{figure}

In order to understand how we can herald the state of the TLS, we perform a running average on the $\{I_i\}$ over $n = 30$ values of $I_i$ defining 
\begin{align}
    \overline{I}^{30}_i = \frac{1}{30}\sum_{k=-15}^{14} I_{i+k}.
\end{align}
The histograms of the $\{\overline{I}^{30}_i\}$ are shown Fig.~\ref{fig:ramsey_corr}c in blue. If we artificially cancel the effect of any correlation in the TLS dynamics on the measurement record $I_i$ by 
shuffling the values $I_i$, we obtain the orange histogram in \ref{fig:ramsey_corr}c. Since both histograms differ, it implies that the measurement records contain extractable information about the TLS.

\begin{figure}[h!]
\begin{center}
  \includegraphics[width=1.0\linewidth]{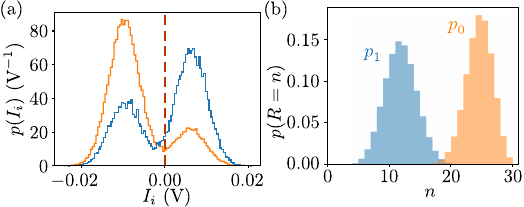}
  \caption{(a) Probability densities of $I_i$ post-selected on the value of $\overline{I}^{30}_i > -3.3~\mathrm{mV} $ (blue) and $\overline{I}^{30}_i \leq -3.3~\mathrm{mV} $.(b) Estimated probability density for the value of $R$ conditionned on the TLS state.} 
  \label{fig:selection_TLS}
 \end{center}
\end{figure}

We plot the histograms of the measurement record $I_i$ post-selected on the condition $\overline{I}^{30}_i>-3.3$ mV (Fig.~\ref{fig:selection_TLS}a blue histogram) and $\overline{I}^{30}_i<-3.3$ mV (Fig.~\ref{fig:selection_TLS}a orange histogram). The value of the threshold is indicated in dashed line Fig.~\ref{fig:ramsey_corr}c.

We can now define the measurement records
\begin{align}
r_i= \left\{
    \begin{aligned}\label{record}
        &1 \mathrm{~if~I_i<0~V}\\
        &0 \mathrm{~~else} 
    \end{aligned}
\right.
\end{align}
and infer a probability $p(r_i = 1|\mathrm{TLS~in~ground~state}) = p_0 = 0.80 $ and $p(r_i = 1|\mathrm{TLS~excited}) = p_1 = 0.45$. The inferred corresponding conditional probability laws of $R$ $p(R=n|\mathrm{TLS~excited})$ and $p(R=n|\mathrm{TLS~in~ground~state})$ are binomial laws with parameters 30 and $p_{0/1}$. Histograms of these laws are shown in orange (for $p_0$) and blue (for $p_1$) in Fig.~\ref{fig:selection_TLS}b. We deduce from this histogram that $p(R\geq 20|\mathrm{TLS~excited}) = 0.0013$, while $p(R\geq 20|\mathrm{TLS~in~ground~state}) = 0.975$, justifying the heralding threshold presented in Fig.~\ref{fig:res_ramsey}a.

Note that the heralding fidelity is actually a bit lower than this, as $I_i$ enters in the computation of $\overline{I}_i$, which introduces a bias. This bias is sufficiently small so that the heralding procedure still works, as shown in Fig.~\ref{fig:res_ramsey}.

\subsection{Heralding the cavity in the vacuum state}
\label{subsection:vacuum}
\begin{figure*}[t]
\begin{center}
  \includegraphics[width=1.00\linewidth]{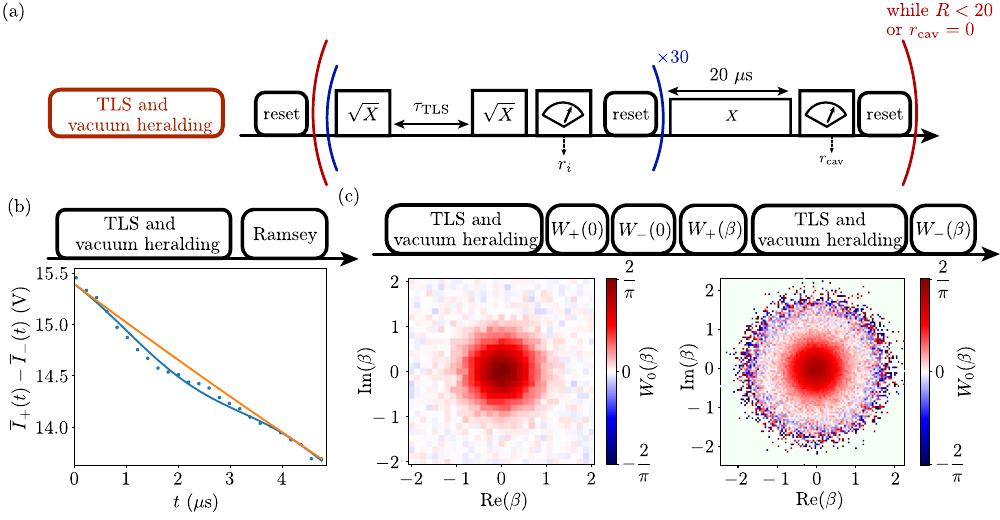}
  \caption{(a) Pulse sequence for the TLS and vacuum heralding procedure. (b) top: Pulse sequence and Ramsey measurement on the qubit. Blue points: measured quadrature. Solid blue line: fit with a thermal population of 0.6~\% using Eq.~(\ref{eq:popRamsey}). Orange: expected signal for a pure vacuum. (c) top: high-level pulse sequence for the measurement of the Wigner of the vacuum. Bottom left: Wigner measurement with square uniform sampling. Bottom right: Wigner measurement with optimal sampling.}
  \label{fig:vac_prep}
 \end{center}
\end{figure*}

The cavity is reset to the vacuum state before each State Preparation and Measurement (SPM) sequence. The complete sequence thus contains an upgraded version of the TLS heralding sequence that includes a Gaussian $20~\mathrm{\mu s}$ long selective $\pi$ pulse on the qubit at $\omega_{\mathrm{q}}/2\pi - 0\chi/2\pi$ at the end (see Fig.~\ref{fig:vac_prep}a). The corresponding linewidth of this truncated Gaussian pulse is $50$ kHz, well below $\chi/2\pi$, which ensures the selectivity of this gate. The readout of the qubit gives $r_\mathrm{cav}= 1$ if the recorded quadrature goes below the threshold (dashed yellow in Fig.~\ref{fig:RO}). The full heralding sequence then consists in a while loop that exits when both the cavity and the TLS are found in their ground state. 


In order to test the performance of this procedure in resetting the cavity population, a Ramsey measurement is performed after each heralding sequence. The result is shown in Fig.~\ref{fig:vac_prep}b. When the cavity is occupied by a thermal state with $n_\mathrm{th}$ average photons, the Ramsey signal is supposed to be proportionnal to $e^{-t/T_2}(1-n_\mathrm{th}+ n_\mathrm{th}\cos{\chi t})$. In blue, the measured signal and fit show a cavity population of $n_\mathrm{th}=0.6~\%$ and fitted $T_2$ is here 40~$\mathrm{\mu s }$ after the heralding procedure, to be compared to the orange curve, corresponding to the same fit but with the cavity rigorously in the vacuum. This average photon number $n_\mathrm{th}$ is attributed to the imperfect QNDness of the qubit readout. 

We also measure the Wigner function $W_0$ of the cavity after this reset Fig.~\ref{fig:vac_prep}c with a SPM sequence showed Fig~\ref{fig:SPM}(a), without playing the NN pulses. The left one is a Wigner tomography with $31\times 31$ pixels averaged 1000 times each, and gives a fidelity $\mathcal{F}^\mathrm{square}_0 = 99.4~\%\pm0.5~\%$ to the vacuum. The right one is a measurement of $2\hat{\Pi}(\beta)/\pi$ using the optimal sampling $p_\mathrm{opt}$ detailed in the text.  The estimated fidelity is $\mathcal{F}^\mathrm{opt}_0 = 98.9~\%\pm 0.15~\%$. We attribute the apparent disagreement with the experiment to a remaining population of the qubit after the reset possibily due to a slightly destructive behaviour of the readout.

\subsection{Pulse calibration}\label{pulse calibration}

The control pulses consist in two pairs of voltages signals $(I_{\mathrm{c}}(t)$, $Q_{\mathrm{c}}(t))$ and $(I_{\mathrm{q}}(t), Q_{\mathrm{q}}(t))$ each upconverted by the OPX controller and a mixer so that the drive amplitudes in the Hamiltonian read $\varepsilon_{\mathrm{c}}(t) = \xi_{\mathrm{c}}(I_{\mathrm{c}}(t)+i Q_{\mathrm{c}}(t))$  and $\varepsilon_{\mathrm{q}}(t) = \xi_{\mathrm{q}}(I_{\mathrm{q}}(t)+i Q_{\mathrm{q}}(t))$. Determining $\xi_{\mathrm{q}}$ and $\xi_{\mathrm{c}}$ is required for a proper implementation of the pulse sequences generated by the NN. 

\begin{figure}[h!]
\begin{center}
  \includegraphics[width=\linewidth]{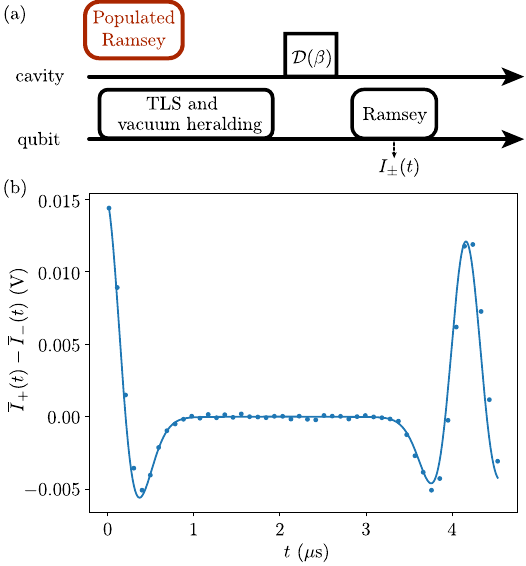}
  \caption{(a) Pulse sequence of the populated Ramsey measurement scheme. (b) Blue dots: measured signal for a $200$ ns long displacement pulse of amplitudes $I_{\mathrm{c}}(t) = 0.03$ V and $Q_{\mathrm{c}}(t)=0$. Solid line: fit using Eq.~(\ref{eq:popRamsey}) with $T_2=17.5~\mathrm{\mu s}$ and $\beta = 2.1$.}
  \label{fig:populated_ramsey}
 \end{center}
\end{figure}

Calibrating $\xi_{\mathrm{q}}$ is done using a standard measurement of Rabi oscillations. Calibrating $\xi_{\mathrm{c}}$ is done using a so-called populated Ramsey sequence, which is a Ramsey sequence at resonance with the qubit, which is performed after a displacement of amplitude $\beta$ on the cavity (see \ref{fig:populated_ramsey}a). The measured Ramsey signal reads \cite{bretheau2015quantum}
\begin{align}
\begin{split}
    \overline{I}_+(t)-\overline{I}_-(t)\propto &\exp[{-\frac{t}{T_2}}]\\
    \times&\exp[|\beta|^2(\cos(\chi t)-1)]\cos(|\beta|^2\sin(\chi t)).\label{eq:popRamsey}
\end{split}
\end{align}
The curve Fig.~\ref{fig:populated_ramsey}b fitted with this expression indicates a displacement amplitude $\beta = 2.1$ after a $200$ ns long displacement pulse of amplitudes $I_{\mathrm{c}}(t) = 0.03$ V and $Q_{\mathrm{c}}(t)=0$ which gives $\xi_{\mathrm{c}}/2\pi = 55.2~\mathrm{MHz/V}$, as well as a proper measurement for $\chi$. The fitted coherence time in the presence of cavity occupation is $T_2 = 17.5~\mathrm{\mu s}$. What limits the $T_2$ in this case is the dephasing induced by the population of the cavity.

\section{State preparation and fidelity measurement}

\subsection{Fidelity estimation}
\label{sec:fidelity_estimation}

The goal of this section is to define the fidelity estimators used to construct the Figure~\ref{fig:fidelities} of the main text. For any target state  $\ket{\mathcal{T}} = \ket{\psi_{\mathrm{c}}} \ket{0}$, the fidelity of the prepared state $\rhohat$ of the cavity and qubit bipartite system is defined as
\begin{align}
    \mathcal{F} = \langle\mathcal{T}|\rhohat|\mathcal{T}\rangle.
\end{align}
Using the decomposition $\rhohat = \rhohat^c_{00}\ketbra{0}{0}+ \rhohat^c_{11}\ketbra{1}{1} +  \rhohat^c_{01}\ketbra{0}{1}+\rhohat^c_{10}\ketbra{1}{0}$, we have
\begin{align}
    \mathcal{F} = \langle\psi_{\mathrm{c}}|\rhohat^c_{00}|\psi_{\mathrm{c}}\rangle = p_0 \langle\psi_{\mathrm{c}}|\rhohat^c_{0}|\psi_{\mathrm{c}}\rangle,
\end{align}
where $\rhohat^c_{0} = \frac{\rhohat^c_{00}}{\mathrm{Tr}({\rhohat^c_{00}})}$ is a properly normalized density matrix for the cavity, and $p_0 =\mathrm{Tr}({\rhohat^c_{00}}) $ is the probability of finding the qubit in the ground state. This fidelity is thus the product of the probability of finding the qubit in the ground state and the fidelity of the cavity conditioned to the qubit being in its ground state:
\begin{align}
    \mathcal{F} = p_0\mathcal{F}^c.
\end{align}
The probability $p_0$ is directly obtained by measuring the qubit at the end of the pulse sequence. Measuring $\mathcal{F}^c$ can be done with Wigner  tomography: we use the fact that
\begin{align} \label{Wigner fidelity}
    \mathcal{F}^c = \pi \int_\mathbb{C} W_{|\psi_{\mathrm{c}}\rangle\langle \psi_{\mathrm{c}}|}(\beta)W_{\rhohat^c_{0}}(\beta)\mathrm{d}\beta.
\end{align}
where $W_{|\psi_{\mathrm{c}}\rangle\langle \psi_{\mathrm{c}}|}(\beta)$ is the Wigner function of the target state, and $W_{\rhohat^c_{0}}$ that of the state represented by $\rhohat^c_{0}$.

Let us assume that $\beta$ is picked at random in the complex plane following a law $p(\beta)$. We can rewrite this expression of the fidelity as the expectation
\begin{align}
    \mathcal{F}^c =\pi \mathbb{E}\left[\frac{W_{|\psi_{\mathrm{c}}\rangle\langle \psi_{\mathrm{c}}|}(\beta)}{p(\beta)}W_{\rhohat^c_{0}}(\beta)\right].
\end{align}
This expression still holds when replacing the function ${W}_{\rhohat^c_{0}} (\beta)$ by a random variable $\tilde{W}_{\mathrm{exp}} (\beta)$ representing the outcome of the measurement of the observable $2\hat{\Pi}(\beta)/\pi$ with mean value ${W}_{\rhohat^c_{0}}(\beta)$. Experimentally, the measurement record is indeed typically a binary random number in $\{\frac{2}{\pi}, -\frac{2}{\pi}\}$. A widely used choice for $p(\beta)$ is a square uniform distribution
\begin{align}
p_\mathrm{square}(\beta)= \left\{
    \begin{aligned}
        &\frac{1}{\Delta x\Delta y}~&&\mathrm{if~|Re}(\beta)|\leq \frac{\Delta x}{2}\\
        & &&\mathrm{and~|Im}(\beta)|\leq \frac{\Delta y}{2}\\
        &0 && \mathrm{otherwise} 
    \end{aligned}
\right.
\end{align}
approximated by averaging $\tilde{W}_{\mathrm{exp}} (\beta)$ on a grid of pixels. In this case, for each $\beta$, $W_{\mathrm{exp}}(\beta)$ is the mean value of $\tilde{W}_{\mathrm{exp}} (\beta_i)$. We choose $n_x n_y$ evenly spaced pixels on a grid of size $\Delta x\times\Delta y$ centered around 0 in the complex plane. This is the most natural way of reconstructing the Wigner function, as it is visually exhaustive. However, it is not the most efficient in terms of the number of samples. In~\cite{Sivak2022}, it is demonstrated that the most efficient sampling strategy is given by
\begin{align}
    p_\mathrm{opt}(\beta) = \frac{|W_{|\psi_{\mathrm{c}}\rangle\langle \psi_{\mathrm{c}}|}(\beta)|}{\parallel W_{|\psi_{\mathrm{c}}\rangle\langle \psi_{\mathrm{c}}|} \parallel_1}
\end{align}
with $\parallel W_{|\psi_{\mathrm{c}}\rangle\langle \psi_{\mathrm{c}}|} \parallel_1 = \int_\mathbb{C} |W_{|\psi_{\mathrm{c}}\rangle\langle \psi_{\mathrm{c}}|}(\beta)|\mathrm{d}\beta$.
The procedure then consists in sampling $N$ points $\{\beta_i\}$ in the complex plane according to the law $p_\mathrm{opt}$, and in recording one value $\tilde{W}_\mathrm{exp} (\beta_i)$ for each of these points. We can then build the estimators $\tilde{\mathcal{F}} = p_0 \tilde{\mathcal{F}}^c$ of fidelity for both strategies. In general, we have
\begin{equation}
     \tilde{\mathcal{F}^c}=\frac{\pi}{N} \sum_i \frac{W_{|\psi_{\mathrm{c}}\rangle\langle \psi_{\mathrm{c}}|}(\beta_i)}{p(\beta_i)}\tilde{W}_\mathrm{exp}(\beta_i).
\end{equation}
Therefore, in the case of the uniform sampling, the fidelity estimator reads
\begin{equation}
 \tilde{\mathcal{F}^c}_\mathrm{square} =\frac{\pi\Delta x \Delta y}{N} \sum_i W_{|\psi_{\mathrm{c}}\rangle\langle \psi_{\mathrm{c}}|}(\beta_i)\tilde{W}_\mathrm{exp}(\beta_i).\label{eq:Fsquare}
\end{equation}
In contrast, in the case of the optimal sampling it reads
\begin{equation}
\tilde{\mathcal{F}^c}_\mathrm{opt} =\frac{\pi\parallel W_{|\psi_{\mathrm{c}}\rangle\langle \psi_{\mathrm{c}}|} \parallel_1}{N} \sum_i\frac{W_{|\psi_{\mathrm{c}}\rangle\langle \psi_{\mathrm{c}}|}(\beta_i)}{|W_{|\psi_{\mathrm{c}}\rangle\langle \psi_{\mathrm{c}}|}(\beta_i)|}\tilde{W}_\mathrm{exp}(\beta_i).
\end{equation}
   
\begin{figure}[ht]
\begin{center}
  \includegraphics[width=1.00\linewidth]{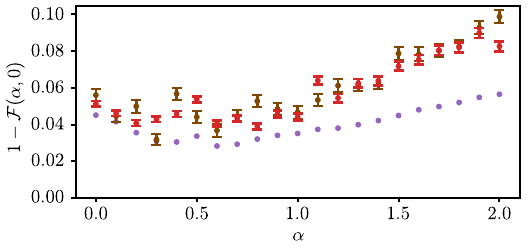}
  \caption{Red dots: Fidelity measured with optimal sampling. Brown dots: Fidelity measured with square sampling. The error bars are wider for about the same number of samples. Purple dots: theoretical expectation from the solid line in Fig.~\ref{fig:fidelities}a.}
  \label{fig:square_vs_opt}
 \end{center}
\end{figure}

Note that the expression (\ref{eq:Fsquare}) of $\tilde{\mathcal{F}}_\mathrm{square}$ is formally equivalent to computing a discrete version of Eq.~(\ref{Wigner fidelity}) between the target Wigner function and the experimental one. The estimated fidelities are compared for both strategies in Fig.\ref{fig:square_vs_opt} where the improvement in the statistical error can be seen in the error bars. Some values differ between both methods, which gives a sense of the  uncertainty originating from the low frequency drifts in the contrast $\tilde{c}$ due to drifts in the qubit coherence time $T_2$ (see Sec.~\ref{sec:effect_of_T2}).

\begin{figure*}[ht]
\begin{center}
  \includegraphics[width=1.00\linewidth]{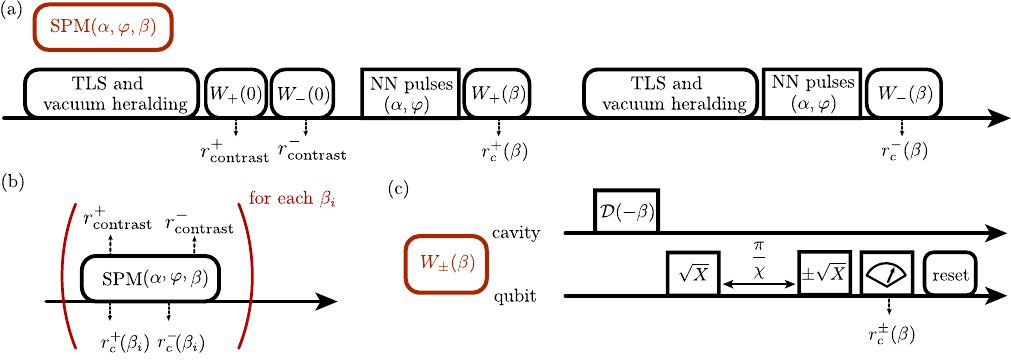}
  \caption{(a) High-level diagram of the SPM sequence. It lasts 8~ms for each realization, which is dominated by the cavity reset time. (b) High-level diagram of the Wigner sampling scheme. (c) Pulse sequence for the Wigner measurement}
  \label{fig:SPM}
 \end{center}
\end{figure*}
\subsection{Experimental implementation}
\label{sec:expt_implementation}

The schematics of the State Preparation and Tomography (SPM) pulse sequence is presented in Fig.~\ref{fig:SPM}a and b. Two parity measurements with opposite polarity are performed for each amplitude $\beta_i$. Wigner measurement of the vacuum at $\beta = 0$ are interleaved to calibrate the possibly drifting contrast of the Wigner measurement.

Before each sequence, a heralding sequence on the TLS and on the cavity emptiness is performed to start with a state as pure as possible before applying the pulse sequence. For each $\alpha$ and $\varphi$, the SPM sequence is performed for $N= 10^5$ times for the optimal strategy, and $N = 21\times 51 \times10^2$ times for the square strategy. Knowing that the cavity is on average in a thermal state with $n_\mathrm{th} = 0.006$, we now define the experimental contrast $\tilde{c}$ and  Wigner estimator $\tilde{W}_\mathrm{exp}(\beta_i)$:
\begin{align}
    &\tilde{W}_\mathrm{exp}^\pm(\beta_i) = \frac{2}{\pi}r_c^\pm(\beta_i)\\
    &\tilde{W}_\mathrm{exp}(\beta_i) =  \tilde{W}_\mathrm{exp}^+(\beta_i) -\tilde{W}_\mathrm{exp}^-(\beta_i) 
\end{align}
Note that with this way of sampling the Wigner function, $\tilde{W}_\mathrm{exp}(\beta_i) $ is a random number that can take discrete values in $\{-\frac{2}{\pi}, 0, \frac{2}{\pi}\}$.

We now define the estimator $\tilde{\mathcal{F}}$ of the fidelity ${\mathcal{F}}$: 
\begin{align}
    &\tilde{c} = \frac{1}{(1-2n_\mathrm{th})N}\sum_{i=1}^N (r_\mathrm{contrast, i}^+- r_\mathrm{contrast, i}^-)\label{eq:contrast0}\\
    &\tilde{\mathcal{F}}=\frac{\pi}{\tilde{c}N} \sum_{i} \frac{W_{|\psi_{\mathrm{c}}\rangle\langle \psi_{\mathrm{c}}|}(\beta_i)}{p(\beta_i)}\tilde{W}_\mathrm{exp}(\beta_i)
\end{align} 

This renormalization by the contrast $\tilde{c}$ is necessary to take into account imperfections of the measurement of $\hat{\Pi}(\beta)$, in particular the finite coherence time of the qubit and the $\pi/2$ pulses and readout fidelities. It is also crucial to interleave the contrasts measurement and the fidelity estimation in a SPM sequence, as the contrast typically drifts by $10~\%$ over several realization of fidelity measurement (typically comprising $N=10^5$ SPM sequences). These drifts are the main reason why Fig.~\ref{fig:square_vs_opt} red and brown do not always agree. The contrast $\tilde{c}$ typically ranges from $0.8$ to $0.9$.

\begin{figure}[h!]
\begin{center}
  \includegraphics[width=1.00\linewidth]{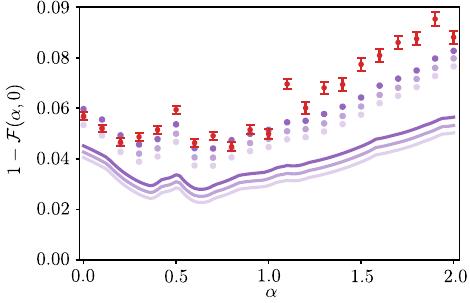}
  \caption{(a) Red dots: measured infidelity $1-\mathcal{F}(\alpha,0)$. Each point is obtained by averaging $10^6$ samples. Purple solid line: simulated infidelity $1-\mathcal{F}(\alpha,0)$ for the same parameters, for $T_2 =42~\mathrm{\mu s}$, $T_2 =50~\mathrm{\mu s}$ and $T_2 =60~\mathrm{\mu s}$ (dark to bright). Purple points: infidelities obtained by simulating also the experimental fidelity measurement procedure and taking into account the finite thermal population of the cavity, for $T_2 =42~\mathrm{\mu s}$, $T_2 =50~\mathrm{\mu s}$ and $T_2 =60~\mathrm{\mu s}$ (dark to bright).}
  \label{fig:fidelities_T2}
 \end{center}
\end{figure}

Since we do not measure the qubit at the end of the NN drive pulses, and do not reset it, doing the Wigner tomography right away biases the estimator $\tilde{\mathcal{F}}$ towards 
\begin{align}
    \mathcal{F}' = \langle\psi_{\mathrm{c}}|\rhohat^c_{00}|\psi_{\mathrm{c}}\rangle - \langle\psi_{\mathrm{c}}|\rhohat^c_{11}|\psi_{\mathrm{c}}\rangle.
\end{align}

\begin{figure*}
\begin{center}
  \includegraphics[width=1.00\linewidth]{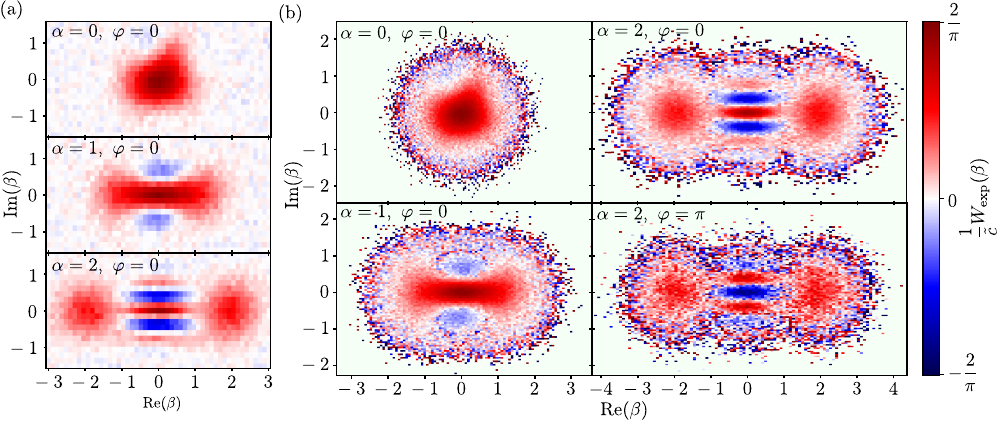}
  \caption{(a) Renormalized reconstructed Wigner functions for several parameters (indicated in the top left) of the neural network, with square sampling. Estimated fidelities are $\mathcal{F}(0, 0) = 93.8~\%\pm0.3~\%$, $\mathcal{F}(2, 0) = 89.5\pm 0.4~\%$. (b) Reconstructed Wigner functions for several parameters of the neural network, with the optimal sampling. $N= 10^6$ samples were used for $\varphi = 0$, $N = 10^5$ for $\varphi = \pi$. estimated fidelities are $\mathcal{F}(0, 0) = 94.3~\%\pm0.2~\%$, $\mathcal{F}(1, 0) = 95.0~\%\pm0.2\pm 0.4~\%$, $\mathcal{F}(2, 0) = 91.1~\%\pm0.3~\%$ and  $\mathcal{F}(2, \pi)= 91.5\pm0.8~\%  $.}
  \label{fig:Wigners}
 \end{center}
\end{figure*}

This can be simply understood: if we post-select the qubit to be in $\ket{1}$ at the beginning of the Wigner pulse sequence, then the probabilities for it to be in $\ket{0}$ or $\ket{1}$ at the end are flipped compared to the situation where it starts in $\ket{0}$. The signal  $\tilde{W}_\mathrm{exp}(\beta_i)$ becomes an estimator for $-W_{\rhohat_1^c}$, where we defined similarly  $\rhohat^c_{1} = \frac{\rhohat^c_{1}}{\mathrm{Tr}({\rhohat^c_{1}})}$.
In the end, our estimator $\tilde{W}_\mathrm{exp}(\beta_i)$ becomes such that:
\begin{align}
    &\mathbb{E}[\tilde{W}_\mathrm{exp}(\beta_i)] = p_0 W_{\rhohat_0^c} (\beta_i)-(1-p_0) W_{\rhohat_1^c}(\beta_i)\\
    &\mathbb{E}[\tilde{\mathcal{F}}] = \mathcal{F}' =\mathcal{F}-\mathcal{F}_1,
\end{align}
where $\mathcal{F}_1 = \bra{\psi_t}\rhohat^c_{1} (1-p_0)\ket{\psi_t}$.

The bias of this estimator thus depends on whether $\rhohat^c_{1} $ is close to $\ket{\psi_{\mathrm{c}}}$ or not.

\subsection{Wigner functions of various states}


A series of additional experimental tomography of several states   is shown Fig~\ref{fig:Wigners}. For the square sampling strategy (Fig.~\ref{fig:Wigners}a), each pixel is averaged 1000 times. To reconstruct Wigner functions from the optimal strategy (Fig.~\ref{fig:Wigners}b), we build an histogram of the $\{\beta_i\}$ with $100\times100$ bins and average the values of the $\{\tilde{W}_\mathrm{exp}(\beta_i)\}$ corresponding to each bin to get ${W}_\mathrm{exp}(\beta_i)$ inside each bin. The resulting Wigner is then very noisy when $W_{|\psi_{\mathrm{c}}\rangle\langle \psi_{\mathrm{c}}|}$ is close to zero. The large Signal-to-Noise Ratio (SNR) thus concentrates on the regions where the target Wigner function has the highest values.

In order to more closely visualize the fidelity of this Wigner function, we renormalize it by $\tilde{c}$. For visualization purposes, we kept the colorbar between $-\frac{2}{\pi}$ and $\frac{2}{\pi}$, even though some pixels contain the average of only a few $\{\tilde{W}_\mathrm{exp}(\beta_i)\}/\tilde{c}\in \{-\frac{2}{\pi\tilde{c}}, 0, \frac{2}{\pi\tilde{c}}\}$, which can be out of range. Note that this does not happen on well averaged pixels. This allows to take into account the imperfections of the measurement apparatus to fully recover the contrast of the experimental Wigner function.

\section{Optimization using GRAPE followed by Krotov}
\label{sec:GRAPE}

In this section, we provide details on the numerical optimization of the control pulses using the more traditional technique denominated as GRAPE followed by Krotov~\cite{QuTiP2,Khaneja2005,Goerz2019}, and about the experimental implementation of these pulses.

\subsection{Numerical optimization}
\begin{figure}[h!]
\begin{center}
  \includegraphics[width=1.00\linewidth]{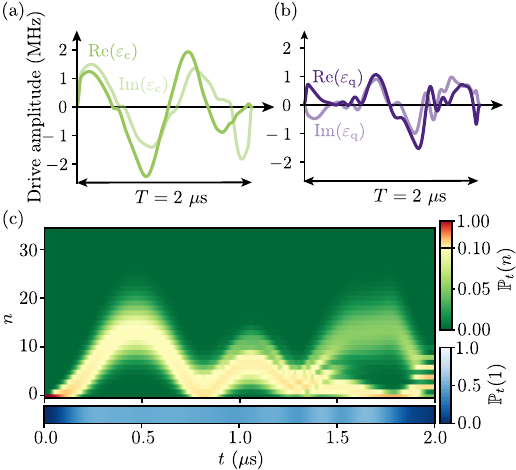}
  \caption{(a,b) Control pulses generated using the GRAPE \& Krotov method for $\alpha = 2$ and $\varphi = 0$.  (c) Simulated probability distribution of the photon number (top) and qubit excitation (bottom) as a function of time for the control fields shown in (a) and (b). The fidelity of the predicted final state to $\Calpha$ is $\mathcal{F}(2, 0) = 95.7~\%$.}
  \label{fig:controlsGRAPE}
 \end{center}
\end{figure}
\begin{figure*}[t]
\centering
\includegraphics[width=0.9\linewidth]{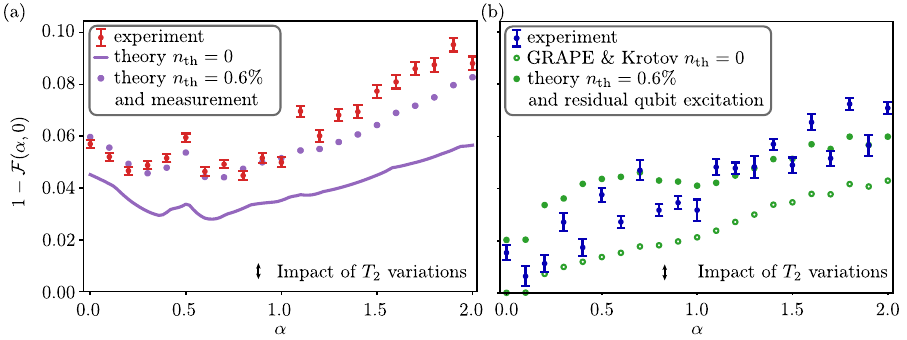}
\caption{(a) Reproduction of Fig.~\ref{fig:fidelities} showing the predicted and measured infidelities when generating the pulses with the NN. (b)  Green open circles: simulated infidelity using pulses optimized by GRAPE followed by the Krotov method~\cite{QuTiP2,Khaneja2005,Goerz2019}. Green dots: infidelities obtained by simulating also the residual qubit excitation during the measurement procedure and taking into account the finite thermal population $n_\mathrm{th}$ of the cavity. Blue dots with error bars: measured infidelity $1-\mathcal{F}(\alpha,0)$ for 21 values of $\alpha$ using the control sequences provided by GRAPE followed by the Krotov method. The error bars are statistical.}
\label{fig:GRAPE}
\end{figure*}

GRAPE and Krotov methods are two optimization algorithms that aim at minimizing the same loss function $\mathcal{L}$, which is the infidelity of Eq.~(\ref{eq:loss_def}). We respectively implemented them in python using the packages QuTiP 4.7.6~\cite{QuTiP2} and Krotov 1.3.0~\cite{Goerz2019}. This implementation of GRAPE only works on Hamiltonian evolution but is easy to parallelize. Feeding the output of GRAPE as an initial guess for the Krotov method allows to take into account decoherence.  In practice, we combine the two for generating the four control sequences $\mathrm{Re}(\varepsilon_{\mathrm{c}}(t))$, $\mathrm{Im}(\varepsilon_{\mathrm{c}}(t))$, $\mathrm{Re}(\varepsilon_{\mathrm{q}}(t))$, and $\mathrm{Im}(\varepsilon_{\mathrm{q}}(t))$ as follows:
    \begin{enumerate}
    	\item  Run GRAPE algorithm based on the unitary evolution set by the Hamiltonian (\ref{H})  on a grid of 163 time steps uniformly spanning a time interval of $1956~\mathrm{ns}$. The algorithm is initialized with sinusoidal functions (one of the few allowed initialization in the optimize\_pulse\_unitary function of control.pulseoptim in QuTiP 4.7.6) ; 
    	\item Upsample the resulting GRAPE optimization results into a denser time grid of 2000 time steps over the total duration time $T=2~\mu\mathrm{s}$. Then smooth the pulses by convolving them with a Gaussian function (standard deviation of 11 ns). This process has the embedded benefit of  flattening the beginning and end of the pulses and to limit the bandwidth of the control signals to an experimentally feasible range;
    	\item Run Krotov's method  based on the Lindblad master equation on the 2000 time steps using the smoothed GRAPE pulses as the initial guess.
    \end{enumerate}

Note that the dimension of the Hilbert space we use grows with cat state amplitude $\alpha$. The dimension $N_\mathrm{max}$ is the nearest integer to $16+4\alpha+2\alpha^2$.

An example of control pulses is shown in Fig.~\ref{fig:controlsGRAPE}a,b for the optimized preparation of $\Calpha$ with $\alpha=2$ and $\varphi=0$. They present more abrupt changes in time than the pulses that are generated by the neural network (Fig.~\ref{fig:pipeline}), which can be explained by the freedom offered by the tuning of every 2000 time steps as opposed to the 9 coefficients of the B-spline decomposition. The photon number distribution is shown in Fig.~\ref{fig:controlsGRAPE}c.

\subsection{Theoretical infidelities}

We solve the Lindblad master equation (\ref{eq:lindblad}) using the solution of GRAPE \& Krotov obtained for each amplitude $\alpha$. The computed infidelity at time $T$ is shown in Figs.~\ref{fig:fidelities}a. 

Since we implement this optimization method experimentally on a different cool-down, we rerun everything for the slightly different parameters of the second cool-down ($\chi/2\pi=237.5~\mathrm{kHz}$, $T_{\mathrm{q}} = 38~\mathrm{\mu s}$, $T_{q, \varphi} = 70.5~\mathrm{\mu s}$ and $T_{\mathrm{c}} = 220~\mathrm{\mu s}$) and plot the predicted fidelities in Fig.~\ref{fig:GRAPE}b as a function of $\alpha$ (green open circles).

To compare with the experimental results, we compute how the estimation of infidelity is affected by the Wigner tomography errors. If we take into account the residual excitation probability of the qubit and the thermal occupation of the cavity $n_\mathrm{th}=0.6~\%$ as in Fig.~\ref{fig:deltaf}, we obtain an overestimation of the infidelity as shown in Fig.~\ref{fig:GRAPE} (green dots). This overestimation is similar in amplitude to the one we find for the control pulses generated by the NN (purple dots compared to purple solid line in Fig.~\ref{fig:fidelities}a).

\subsection{Experimental implementation of GRAPE}
We implement the 21 control sequences generated by GRAPE \& Krotov on a different cool-down for the same device as the one used with the NN. The procedure for the fidelity estimation is the same as the one described in Sec.~\ref{sec:expt_implementation}, except for the computation of $\tilde{c}$. It is now estimated by measuring an evenly distributed grid of $11\times 11$ complex amplitudes $\beta=x+iy$ for the Wigner function $W_0(\beta)$ where $-1\leq x,y\leq 1$, obtained when the cavity is heralded in the vacuum. In contrast, the formula given in Eq.~(\ref{eq:contrast0}) only relies on the point $\beta=0$.

The resulting measured infidelity is plotted as blue dots as a function of $\alpha$ in Fig.~\ref{fig:GRAPE}b. The error bars represent statistical errors. The number of averaging of each point varies owing to frequent temperature rises of the refrigerator triggered by another experiment. The fidelity increases with amplitude $\alpha$ similarly as the prediction. However, qubit frequency drifts and thermal occupation of the cavity varied often during the run leading to a spread of the measured fidelities beyond the statistical uncertainty.

Overall, Fig.~\ref{fig:GRAPE}a,b show that the experimentally observed fidelities are close to the predicted infidelities we should observe, owing to measurement errors. In Fig.~\ref{fig:GRAPE}, the actual state fidelities at the end of the preparation are represented by the solid purple line for the NN and by the open green circles for GRAPE \& Krotov. Both differ by about 0.01 for large amplitudes $\alpha$.

\section{Error analysis}
\label{sec:errors}
\subsection{Error budget}

\begin{figure}
\begin{center}
  \includegraphics[width=1.00\linewidth]{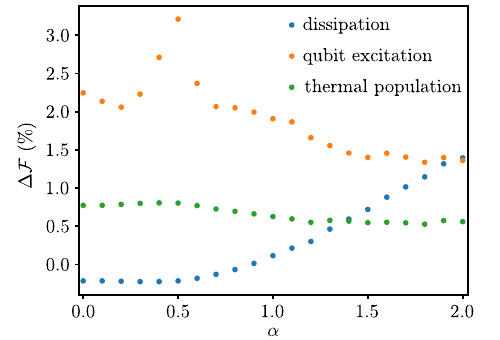}
  \caption{Blue dots: contribution to the infidelity $\Delta \mathcal{F}_\mathrm{dissip}$ due to the dissipation during the parity measurement. Orange dots: contribution due to the residual qubit excitation at the end of the pulse sequence. Green dots: effect of the thermal population of the cavity ${n_\mathrm{th}} = 0.6~\%$.}
  \label{fig:deltaf}
 \end{center}
\end{figure}

In order to understand the difference between the prediction of the state preparation fidelities (purple in Fig.~\ref{fig:fidelities}a) and the measured ones (red in Fig.~\ref{fig:fidelities}a), we numerically solve a Lindblad master equation of the whole measurement sequence that takes into account the three phenomena leading to infidelity losses $\Delta \mathcal{F}_\mathrm{th}$ due to the remaining thermal population of the cavity before the pulse sequence, $\Delta \mathcal{F}_\mathrm{dissip}$ due to dissipation during the parity measurement sequence, and $\Delta \mathcal{F}_\mathrm{ex}$ due to the remaining qubit excited population before the parity measurement sequence. Simulations removing separately these three effects reveal their respective contributions. Results are shown Fig.~\ref{fig:deltaf}. The thermal population $n_\mathrm{th}$ systematically leads to $\Delta \mathcal{F}_\mathrm{th} \simeq n_\mathrm{th} = 0.6~\%$, and $\Delta \mathcal{F}_\mathrm{par}$ varies monotonically from 0.6~\% for small cat states to $2~\%$ for the largest values. It is negative for small alphas, as relaxation towards the vacuum actually improves the fidelity. $\Delta \mathcal{F}_\mathrm{ex}$ is of the order of 2~\%. Its non trivial variations depend on the way the NN was trained.

\subsection{Effect of $T_2$ fluctuations}
\label{sec:effect_of_T2}

The effect of $T_2$ fluctuations are shown in Fig.~\ref{fig:fidelities_T2}. Simulations are run in the same way as Fig.~\ref{fig:fidelities}(a), but $T_2$ of 42 (dark), 50 (medium) and 60 $\mathrm{\mu s}$ (bright) are considered. The mismatch between theory and experiment could thus be explained by variations in $T_2$. However, we still see that the experiment has a tendency to perform worse for larger alphas compared to the prediction. We attribute this either to imprecision in the measured value of $\chi$, or higher order processes taking place during the preparation. The consequences of these two reasons become more important when the cavity is loaded with more photons during the preparation. As the intermediate number of photons in the cavity  typically increases for larger values of $\alpha$, we expect a larger deviation for larger $\alpha$.

\end{document}